\documentclass[aps,prb,nofootinbib,showkeys,showpacs,amsmath,amssymb,amsfonts,twocolumn,floatfix]{revtex4}
\pdfoutput=1
\usepackage{graphicx}
\usepackage[tight,normal]{subfigure}
\usepackage{units}
\usepackage{ae}
\usepackage{hyperref}

\begin{document}
\title{Non-Equilibrium and Quantum Coherent Phenomena in the Electromechanics of Suspended Nanowires} 
\author{Robert I. Shekhter}
\email{robert.shekhter@physics.gu.se}
\affiliation{Department of Physics, University of Gothenburg, SE-412 96 G\"oteborg, Sweden}
\author{Fabio Santandrea} 
\affiliation{Department of Physics, University of Gothenburg, SE-412 96 G\"oteborg, Sweden} 
\author{Gustav Sonne}
\affiliation{Department of Physics, University of Gothenburg, SE-412 96 G\"oteborg, Sweden} 
\author{Leonid Y. Gorelik} 
\affiliation{Department of Applied Physics, Chalmers University of Technology, SE-412 96 G\"oteborg, Sweden} 
\author{Mats Jonson\footnote{Also at the Division of Quantum Phases and Devices, School of Physics, Konkuk University, Seoul 143-107, Korea}} 
\affiliation{Department of Physics, University of Gothenburg, SE-412 96 G\"oteborg, Sweden}
\affiliation{School of Engineering and Physical Sciences, Heriot-Watt University, Edinburgh EH14 4AS, Scotland, UK}
\date{\today}

\begin{abstract}
Strong coupling between electronic and mechanical degrees of freedom
is a basic requirement for the operation of any nanoelectromechanical
device. In this Review we consider such devices and in particular
investigate the properties of small tunnel-junction nanostructures
that contain a movable element in the form of a suspended nanowire. In
these systems, electrical current and charge can be concentrated to
small spatial volumes resulting in strong coupling between the
mechanics and the charge transport. As a result, a variety of
mesoscopic phenomena appear, which can be used for the transduction of
electrical currents into mechanical operation. Here we will in
particular consider nanoelectromechanical dynamics far from
equilibrium and the effect of quantum coherence in both the electronic
and mechanical degrees of freedom in the context of both normal and
superconducting nanostructures.
\end{abstract}

\maketitle

\tableofcontents

\section{\label{sec:Firstintro} Introduction}

\noindent One of the main goals of contemporary nanophysics is to
attain the means for controlled operation of mechanical devices on the
nanometer length scale. The required coupling between electronic and
mechanical degrees of freedom can be dramatically enlarged in
conducting nanoelectromechanical systems (NEMS) where electrical
currents and accumulated charge can be concentrated to small spatial
volumes. Of particular interest in this Review are small tunnel
structures that incorporate a movable element in the form of a
suspended nanowire. This is because here, as we shall demonstrate
below, one is able to harvest a host of mesoscopic phenomena for the
purpose of transduction of electrical currents into nanomechanical
operations.

One example of a nanoelectromechanical (NEM) device is the
nanoelectromechanical single-electron tunneling (NEM-SET)
transistor. Such a device may be built by self-assembly, employing
mechanically soft organic molecules anchored at metal electrodes while
attached to a small conducting "Coulomb dot" or "single-electron box"
\cite{Andres1996,Soldatov1996}. In this way mechanical displacements
of the dot with respect to the conducting leads are possible (see
also Ref.~[\onlinecite{Park2000}]). Other approaches to building
NEM-SET devices involve the use of a flexible nanopillar placed
between a source and drain electrode \cite{Scheible2004} and placing a
metal "dot" on the tip of an externally driven cantilever
\cite{Erbe2001}. Generally, in NEM-SET devices the transport of
electrons is due to a combination of single-electron tunneling events
(between either the source or the drain lead and the dot) and the
mechanical transportation of charge through to the motion of the dot.

Mechanical vibrations can be coupled to the electron transport both
via uncompensated charge (electrostatic coupling) and currents
(magnetomotive coupling). An electrostatic coupling can be realized in
devices where a nanomechanical resonator is used as a movable gate
electrode \cite{LaHaye2004,Naik2006,Knobel2003}. In such
nanoelectromechanical devices, where the movable part is effectively
``zero-dimensional", one has found that the electromechanical coupling
can --- under certain conditions --- have much more dramatic results
than simply vibron assisted tunneling, which follows from a
straightforward application of perturbation theory. In particular, it
has been shown that an electrostatic coupling of mechanical and
electronic degrees of freedom can induce an electromechanical
instability, resulting in self-sustained temporal oscillations in both
the mechanical and electronic characteristics of the system. This
instability is behind the new ``shuttle" mechanism for mechanically
promoted single-electron mesoscopic charge transport that was proposed
in Ref.~[\onlinecite{Gorelik1998}]. A number of the consequences of
mechanically assisted ``shuttle" transport have been investigated for
normal, superconducting and magnetic NEM-SET devices (see, e.g., the
reviews Ref.~[\onlinecite{Shekhter2003}] and
[\onlinecite{Shekhter2007}]).

Other frequently studied nanoelectromechanical systems are
``one-dimensional" in the sense that they incorporate a suspended
nanowire as a (possibly) vibrating current carrying element. In this
case there are many more relevant mechanical degrees of freedom than
in the rigid "zero-dimensional" quantum-dot-type NEM devices. In the
latter type of device it is sufficient to take the
center-of-mass-motion of the movable element into account, while in
the former also flexural and other types of wire vibrations, as well
as intrinsic phonon modes, can couple to the electrons
\cite{Flensberg2006,Sapmaz2003,Poot2007}. The effect of the various
vibration modes on electron transport can be investigated either
through the driven resonator method, where a down-mixing technique is
used for detection \cite{Sazonova2004,Witkamp2006}, or through
scanning tunneling spectroscopy methods, detecting phonon-assisted
channels for the inelastic tunneling of electrons
\cite{Huttel2008,Sapmaz2006,LeRoy2004}.

A number of novel nanoelectromechanical phenomena become possible as
a result of the elongation of the movable NEM-SET element. The
present authors have, e.g., recently been involved in studies of the
interplay in such devices between quantum coherence in both
electronic and vibrational degrees of freedom. Other investigations
have been related to non-equilibrium transport issues resulting from
electromechanical instabilities that drive the system far away
from equilibrium. This work will be reviewed in the present paper,
where the following topics will be discussed.

\begin{itemize}

\item[1.]  {\it Multi-mode shuttle structures}.
The nanoelectromechanics of nanowire-based (``one-dimensional")
NEM-SET devices have to deal with the multimode flexural vibrations
of the suspended wire, and their effect on charge transport. The
relevant theoretical framework will be reviewed, and in particular
we will show how a strong nonlinear coupling between different modes
leads to a ``self organization" effect in multimode shuttle
structures.

\item[2.] {\it Interplay between quantum coherent mechanical
  vibrations and coherent electron transport}. Quantum coherence is
  expected to play a significant role in the nanomechanics of
  sufficiently long suspended carbon nanowires, since the amplitude of
  their zero-point oscillations is comparatively large. The
  entanglement between quantum coherent electrons and quantum coherent
  mechanical vibrations, induced by an external magnetic field, will
  be shown to qualitatively modify the NEM-assisted electron transport
  in this case. This phenomenon suggests that measuring the
  magnetoresistance may be a way of testing quantum coherence in
  mechanical nanovibrations.

\item[3.]  {\it Nanoelectromechanics of a superconducting weak
  nanowire-link}. A current passing through a suspended wire of
  nanometer-sized cross-section in the presence of a transverse
  magnetic field, gives rise to a strong nanoelectromechanical
  coupling via both the Lorentz force and the electromotive force. We
  show that this coupling qualitatively changes the electrodynamics of
  a suspended nanowire serving as a superconducting weak link. The
  possibility to resonantly pump energy into nanowire vibrations by
  means of an AC Josephson current is explored.

\end{itemize}

The paper is organized as follows. In Section~\ref{sec:intro} we
introduce the reader to the important concepts of this Review and in
particular focus on the effects of charge concentration in
nanoelectromechanical tunneling structures. These ideas are expanded
upon further in Section~\ref{subsec:si_singlemode} and
\ref{subsec:si_multimode} where the conditions for the single- and
multimode shuttle instability respectively are derived. We conclude
this part of the Review by demonstrating how self-selectivity between
the vibrational modes of the considered suspended nanowire device can
be achieved, Section~\ref{subsec:multistability} and
\ref{subsec:STM_displacement}. In Section~\ref{sec:NEM} we instead
consider the effects of current variations coupled to an externally
applied magnetic field and analyze the subsequent electron-vibron
interaction for the case of normal, Section~\ref{sec:Aharonov}, and
superconducting, Section~\ref{sec:Super}, electrodes.

\section{\label{sec:intro} Electromechanical coupling in tunneling nanostructures with charge concentration}

Electron transport in tunneling nanostructures can be strongly
affected by the accumulation of charge in small parts of the
devices. The increase in electrostatic energy corresponding to such
charge accumulation can easily exceed typical thermal energies and
energies available from the voltage source, resulting in ``Coulomb
blockade" of electron tunneling \cite{Shekhter1972}. The Coulomb
forces associated with the uncompensated electric charge can
furthermore, under the right conditions, be large enough to induce
significant mechanical deformation of the movable parts of
nanomechanical devices. The ``electromechanical instability'' or
``shuttle instability'' of the NEM-SET device considered in
Refs.~[\onlinecite{Gorelik1998,Shekhter2003,Shekhter2007}] is a
remarkable example of this phenomenon. In this system
electromechanical coupling is achieved due to the accumulation of
charge on a movable metal island (``dot") and mechanical ``shuttle"
transport of single electrons can be achieved as a result of an
electromechanical instability if a sufficiently large bias voltage is
applied.

\begin{figure}
\center
\subfigure[]{\label{fig:model}\includegraphics[width=0.55\linewidth]{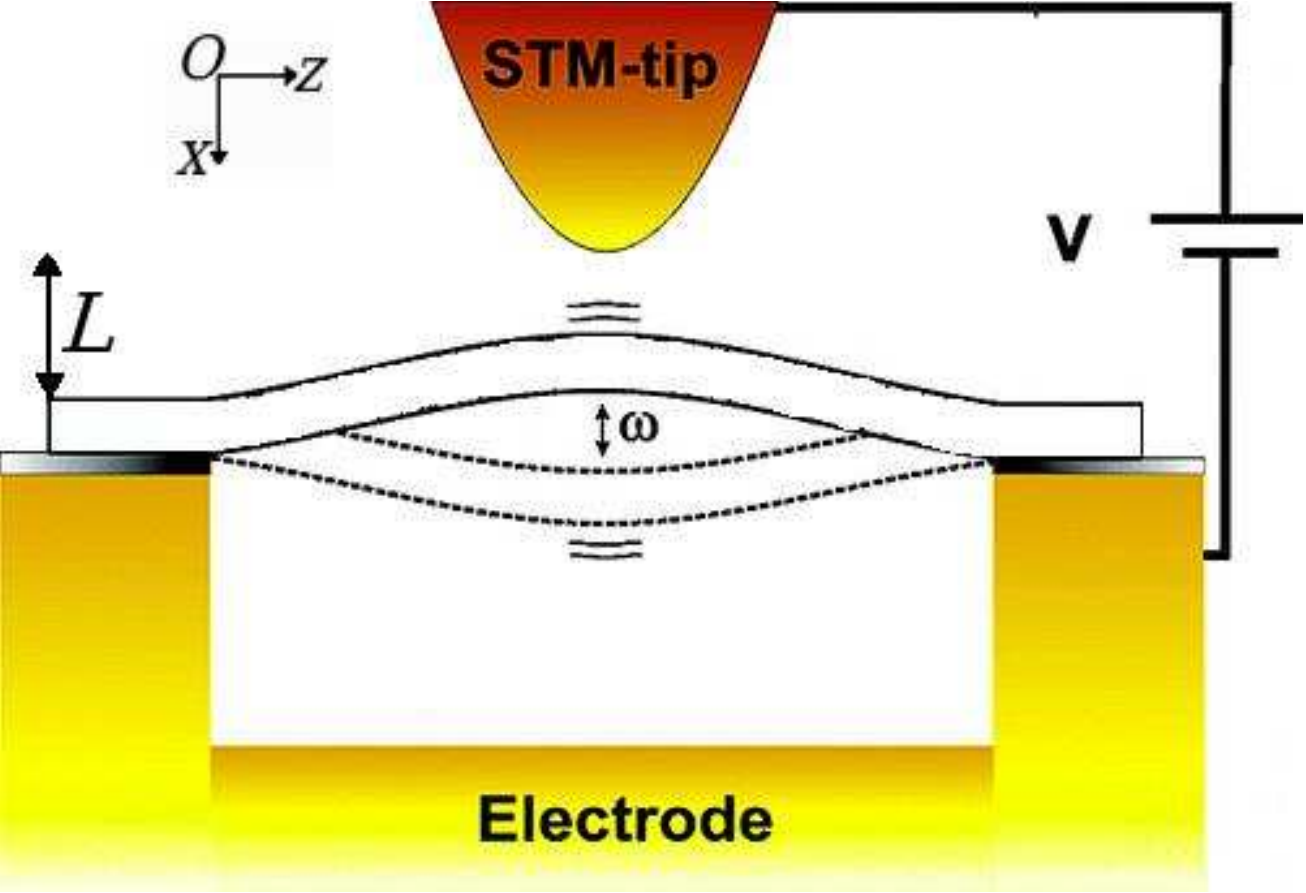}}
\subfigure[]{\label{fig:circuit}\includegraphics[width=0.34\linewidth]{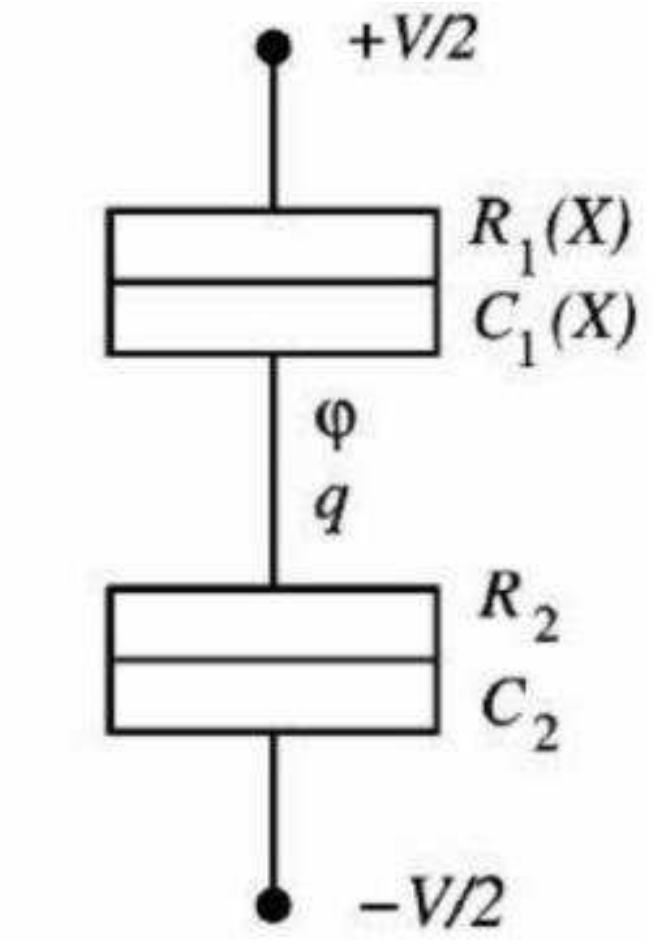}}
\caption{(Color online) Sketch of the system considered \textbf{(a)}
  and the equivalent circuit \textbf{(b)}. Mechanical deviation of the
  wire away from the static configuration is described by the
  ``displacement field'' $u(z,t)$.  An STM tip is put over the point
  $z_0$ along the suspended nanowire axis and the electron tunneling
  rate $\Gamma_1(u(z_0)) = [R_1(u(z_0))C_1(u(z_0))]^{-1}$ between the
  STM tip and the nanowire depends on the spatial profile of the
  nanowire. The tunneling rate $\Gamma_2 = [R_2C_2]^{-1}$ between the
  nanowire and the electrodes is, however, constant.  Adapted with
  permission from [\onlinecite{Jonsson2005}], L. M. Jonsson \textit{et
    al., Nano Lett.}, \textbf{5}, 1165 (2005). $\copyright$ 2005,
  American Chemical Society.}
\end{figure}

In the following sections we will explore the possibility to utilize
the flexural vibrations of a suspended nanowire to transport electric
charge. Such mechanically assisted transport can be attained if the
electric charge injection is focused into the movable part of the
device. This can be accomplished through the use of a scanning tunnel
microscope (STM) which locally injects current into the vibrating
suspended nanowire as shown in Fig.~\ref{fig:model}.

An experimental realization of this device has been studied by LeRoy
\textit{et al.} \cite{LeRoy2004} and details on the fabrication of the
device can be found in Ref.~[\onlinecite{LeRoy2004a}]. In this work,
the resonator consists of a single-wall carbon nanotube (CNT)
suspended over a trench such that a segment of the tube is free to
move in response to external forces. Since the shuttle (in)stability
analysis presented here does not crucially depend on the material of
which the movable part is made, we will in the following consider a
generic oscillating nanowire.

In order to fully describe the mechanics of the nanowire, the complete
set of its flexural vibrational modes should be taken into account. It
will be shown later that, under certain conditions, this adds some new
features to the ordinary picture of nanomechanical shuttling in
``point-like'' movable islands. However, the conditions for the
electromechanical instability to occur are not crucially affected by
the multimode nature of flexural vibrations, hence we will start the
analysis of the shuttle instability considering only a single
vibrational mode.

\subsection{\label{subsec:si_singlemode} Single mode instability}

The device shown in Fig.~\ref{fig:model} models the nanowire as a beam
of length $l$ with clamped ends. Let the undeformed wire extend along
the $z$-axis, while its cross-section lies in the $xy$-plane.  If the
wire is deformed from its static configuration by a bending force
perpendicular to the $z$-axis, it will start to oscillate in the
$xz$-plane.  Such flexural vibrations of CNTs have been detected in
devices similar to the one analyzed here
\cite{LeRoy2004,Sazonova2004,Izumida2005,Witkamp2006,Zazunov2006,Flensberg2006,Sapmaz2006}.

The deformation of the nanotube profile along the $x$-direction can be
described by the displacement field $u(z,t)$, which, according to
linear elasticity theory \cite{Poot2007,Landau1986} obeys the
following equation of motion,
\begin{equation} \label{eq:elast}
\rho S \frac{\partial^2 u}{\partial t^2} + EI \frac{\partial^4 u}{\partial z^4} = f\,.
\end{equation}
Here, $E$ is the Young's modulus of the nanowire, $I$ is the
cross-sectional area moment of inertia, $\rho$ is the linear mass
density, $S$ is the cross-section area and $f$ is the force per unit
length acting on the nanowire. The boundary conditions for the
doubly-clamped beam are $u(0,t)=u(l,t)=0$ and $\partial u /
\partial z(0,t) = \partial u / \partial z(l,t) = 0$.

The displacement field and the force per unit length in
Eq.~\eqref{eq:elast} can be expanded in the complete set of normal
modes $\{\varphi_j(z)\}$ obtained by diagonalization of the hermitian
operator $d^4/dz^4$ with the above boundary conditions. The result is
\begin{subequations} \label{eq:uf_normal_modes}
\begin{align}
u(z,t) & = \sum_j x_j(t) \varphi_j(z)  \\*
f(z,t) & = \sum_j f_j(x_1(t), x_2(t), \ldots, t) \varphi_j(z)\,,
\end{align}
\end{subequations}
where $\{x_j(t)\}$ and $\{f_j(x_1(t), x_2(t), \ldots, t)\}$ are
time-dependent amplitudes.  Note that we write the amplitudes for the
force as $\{f_j(x_1(t), x_2(t), \ldots, t)\}$ instead of just
$\{f_j(t)\}$ to stress that, in the system considered, the force at
any time $t$ depends also on the displacement of the wire, i.e. on the
normal mode amplitudes $\{x_j(t)\}$. Inserting the expressions
\eqref{eq:uf_normal_modes} for $u(z,t)$ and $f(z,t)$ into the equation
of motion \eqref{eq:elast} for the beam, one finds that the beam
dynamics is equivalent to that of a set of coupled harmonic
oscillators $\{x_j\}$ with frequencies $\{\omega_j\}$, affected by
forces $\{f_j(x_1, x_2, \ldots, t)\}$. In general, the unperturbed
modal oscillation frequencies are given by $\omega_j
=(c_j/l^2)(EI/\rho S)^{1/2}$, where the coefficients $c_j = 22.4,
61.7, 120.9, 199.9, 298.6, \ldots$ are obtained by solving the
equation $\cos \sqrt{c_j} \cosh \sqrt{c_j} = 1$, see
Ref.~[\onlinecite{Landau1986}] for details.

The normal mode eigenfunctions $\{\varphi_j(z)\}$ are well
approximated by $\{\sin (j\pi z/l) \}$ and can be classified according
to their symmetry properties; either ``even'' or ``odd'' under the
spatial inversion operation $z \rightarrow -z$ with respect to the
midpoint of the nanotube. The modes labeled $1, 3, \ldots, 2j+1,
\ldots$ turn out to be even, while the modes $2, 4, \ldots, 2j,
\ldots$ are odd. For the moment we assume that the STM tip is put over
the midpoint of the nanowire ($z_0 = l/2$).  This assumption enables
us to neglect the coupling between the STM and the odd
modes. Furthermore, only the fundamental bending mode is taken into
account, for which $\varphi_1(l/2) \sim 1$ (if the normal modes
$\{\varphi_j(z)\}$ are normalized to $1$). The displacement of the
nanowire then reduces to the amplitude of the first mode (for the rest
of this section indicated as $x(t)$ instead of $x_1(t)$).  Finally, a
DC bias voltage $V$ is applied between the STM and the electrodes
enabling electrons to tunnel from the STM tip to the nanowire and then
from the nanowire to the electrodes.

From the point of view of the charge transport the system can be
modeled as two tunnel junctions in series. The junction between the
STM tip and the nanowire, labeled ``1'' in Fig.~\ref{fig:circuit}, is
characterized by a capacitance $C_1(x)$ and a resistance $R_1(x)$,
both of which depend on the deflection of the tube.  This dependence
is assumed to be of the form $R_1(x) \sim e^{-(L+x)/\lambda} \equiv
R_0e^{-x/\lambda}$, where $\lambda$ --- the tunneling length --- is a
parameter used to characterize the tunnel barrier. If the wire
oscillates, its separation from the STM tip changes in time, hence so
does the tunneling probability. The interface between the nanowire and
the metal electrodes is also assumed to be a tunnel junction (rather
than an ohmic contact). It is labeled ``2'' in Fig.~\ref{fig:circuit}
and is characterized by two constants, the capacitance $C_2$ and the
resistance $R_2$.

When an electron tunnels onto the nanowire, a certain time $\tau_q$ is
needed in order to redistribute the charge on it. This time can,
however, always be assumed to be shorter than the other characteristic
times in the system. These are
\begin{itemize}
 \item $\tau_1$, the time needed for electrons to tunnel from the STM
   tip to the nanowire
 \item $\tau_2$, the time needed for electrons to tunnel from the
   nanowire to the electrodes
 \item $T$, the period of nanowire oscillations.
\end{itemize}
Assuming that $\tau_q$ can be neglected allows us to treat the wire as
a metal island with a well-defined excess charge $q$ and an
electrostatic potential $\Phi$.  When the wire is charged by electrons
tunneling from the STM tip, it is affected by a capacitive force
$F_{\rm{cap}}$ whose intensity depends on the voltage drop between the
STM tip and the wire.

The question is now under what circumstances a shuttle instability can
be expected to occur in the system, a problem that was initially
addressed by Jonsson \textit{et al.} \cite{Jonsson2005}. Here, we
remind the reader that the shuttle instability requires that the
static equilibrium state of the wire is unstable with respect to the
formation of a new dynamical stationary state characterized by
finite-amplitude oscillations, so called \textit{limit cycle
  oscillations} \cite{Strogatz2001}, around the static configuration.

To answer the above posed question, let us consider the case where the
nanowire is only slightly perturbed from its stationary configuration
and let free to oscillate. The only thing that can then happen, given
that the bias voltage $V$ is not too large, is that it performs some
damped oscillations and returns to its initial configuration. In
general a broad variety of microscopic mechanisms contribute to such
damping of oscillations in nano-resonators. Some of them, e.g.,
thermoelastic damping \cite{Lifshitz2000}, are related to internal
features of the resonator while others, e.g., losses due to the
clamping \cite{Hao2003} and to air friction \cite{Blom1992}, depend on
its interaction with the environment.  For the moment we assume that
in the system considered all dissipative effects can be taken into
account through a phenomenological viscous damping term $-\gamma
\dot{x}$ (the effects of other types of damping will be considered in
Section~\ref{subsec:STM_displacement}).

If the wire is instead perturbed further and allowed to perform larger
mechanical vibrations the situation is somewhat altered. In this
scenario the wire can be very close to the STM tip ($u \lesssim
\lambda$), in which case we have that $R_1 \ll R_2$, i.e. $\tau_1 \ll
\tau_2$. Under these conditions tunneling from the STM tip is very
likely and the electrostatic potential $\Phi$ of the wire is
approximately the same as that of the STM tip. On the other hand, if
the wire is instead far from the STM tip ($u > \lambda$), the opposite
is true and $\Phi$ approaches the potential of the electrodes. Now, if
the wire oscillates with frequency $\omega$, the parameters $\omega
\tau_1$, $\omega \tau_2$ have some finite values which implies that at
any moment $\Phi$ has no time to adjust itself to the equilibrium
value determined by the wire's position. As a result it also depends
on the wire's position and velocity at earlier times. In other words,
there is a correlation between the wire's velocity and the
electrostatic force. This in turn implies that the force performs some
work on the wire, $\langle \dot{x}F_{\rm cap}\rangle \neq 0$, where
$\langle \ldots \rangle$ is the time average over one period.

According to the ordinary shuttle theory, if the work done by the
electrostatic force overcomes the work done by the dissipative forces,
the electromechanical instability occurs. In order to check these
qualitative considerations, the evolution of the system has been
modeled in Ref.~[\onlinecite{Jonsson2005}] using the following
equations of motion for the position of the wire's midpoint, $x$, and
the excess charge, $q$, on it
\begin{subequations}\label{eq:eom1}
\begin{align}
\ddot{x} + \gamma \dot{x} + \omega x & = \frac{F_0(C_2V-q)^2}{\left( 1-\frac{x}{L} \right)^2} \equiv F_{\rm cap} \\*
\dot{q} & =  \frac{V}{2} \left( G_-(x) - \frac{C_-(x)}{C_{\Sigma}(x)}\right) - q\frac{G_{\Sigma}(x)}{C_{\Sigma}(x)}\,.
\end{align}
\end{subequations}
In Eq.~\eqref{eq:eom1}, $C_{\Sigma}(x) \equiv C_1(x) + C_2$ is the
total capacitance, $C_-(x) \equiv C_1(x)-C_2$, $G_{\Sigma} \equiv
1/R_1(x) + 1/R_2$ is the total conductance and $G_-(x) \equiv
1/R_1(x)-1/R_2$. Due to the short distance between STM tip and
nanowire, surface forces might also be significant. However, their
relevance compared to the elasto-mechanical forces can always be
reduced by a suitable choice of length and thickness of the wire
\cite{Jonsson2005}. The excess charge on the tube, $q$, is treated as
a continuous variable, an approximation which is justified if the
tunneling times $\tau_1$, $\tau_2$ are much smaller than the period of
oscillation and the temperature is not too low.

In order to perform a stability analysis of \eqref{eq:eom1}, one can
linearize the two equations around the stationary solution ($x_0$,
$q_0$). From this analysis it is found that if the dissipation
coefficient exceeds a certain threshold value, $\gamma >
\gamma_{\rm{thr}}$ for a fixed bias voltage $V$, the stationary
solution is a stable fixed point. Therefore, any trajectory that
starts close to this solution will ultimately fall into this point.
However, if instead the dissipation coefficient is smaller than the
threshold value, $\gamma < \gamma_{\rm{thr}}$, any small initial
deviation from ($x_0$, $q_0$) grows in time and the system manifests a
``shuttle-like'' instability. The conditions for this instability can
be reformulated as $1/Q < 1/Q_{\rm{thr}}$, where $Q$ is the ``quality
factor'' of the oscillator. This is an experimentally accessible
parameter that expresses the robustness of the oscillations against
all possible sources of dissipation, hence it can be measured
independently on the specific model used to describe dissipation in
the system. For the case considered here, the $Q$-factor can be
expressed as $Q = 1/\gamma$. Values of $Q$ between $10$ and $10^3$
have been reported for carbon nanotube-based nano-resonators
\cite{Sazonova2004} and of the order of $10^5$ for SiC nanowire-based
oscillators \cite{Perisanu2007}.
\begin{figure}
\centering
\includegraphics[width=0.8\linewidth]{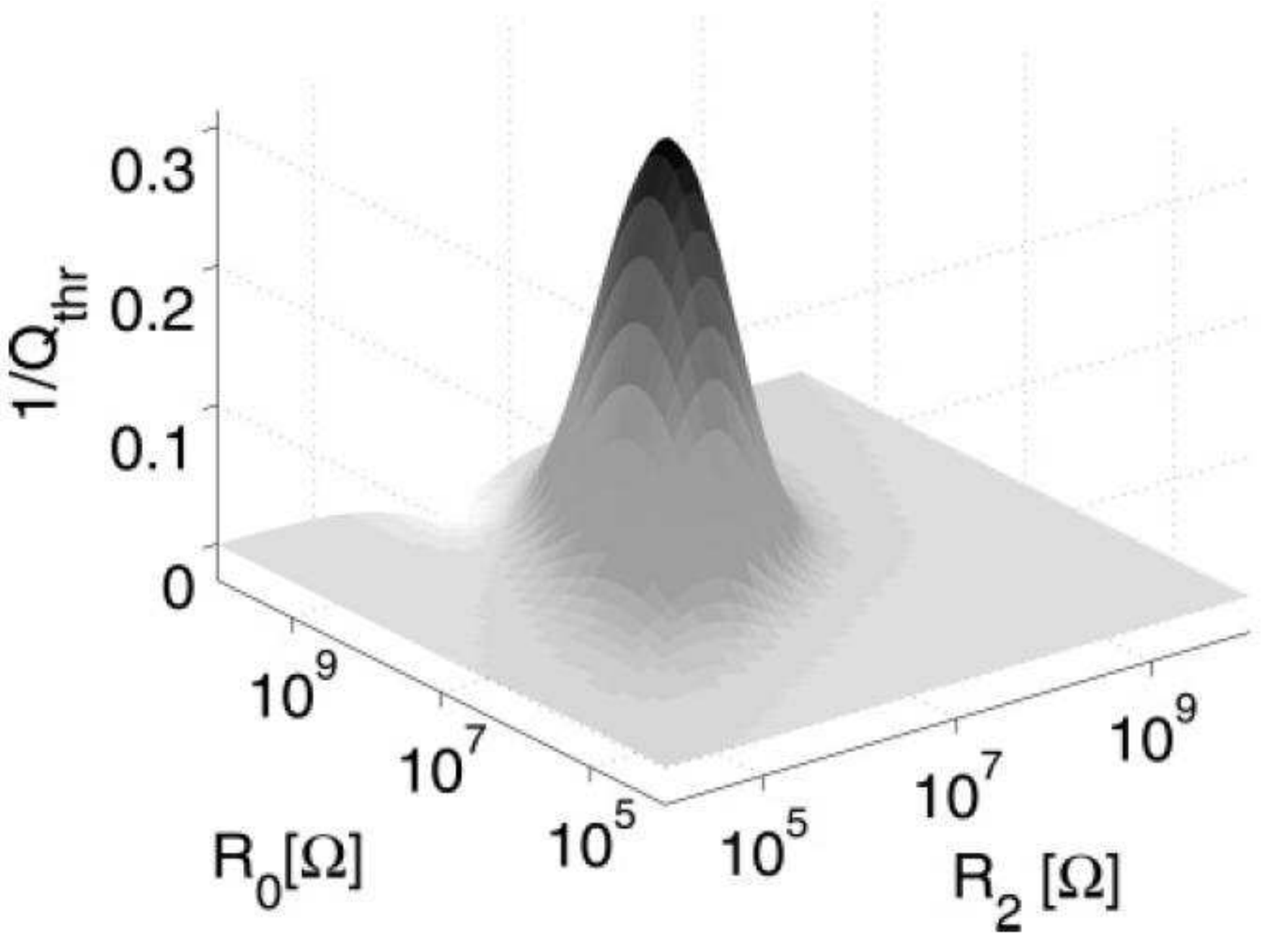}
\caption{Threshold dissipation as a function of the junction
  resistances ({\em Cf}. Fig.~\ref{fig:circuit}). The largest values
  of $1/Q_{\textrm{thr}}$ are obtained when the two junction
  resistances are of the same order of magnitude. Reprinted with
  permission from [\onlinecite{Jonsson2005}], L. M. Jonsson \textit{et
    al., Nano Lett.}, \textbf{5}, 1165 (2005). $\copyright$ 2005,
  American Chemical Society.}
\label{fig:gamma_thr}
\end{figure}

In Fig.~\ref{fig:gamma_thr} we show the threshold dissipation
coefficient $\gamma_{\rm{thr}}$ as a function of the parameters $R_0$,
$R_2$ (for a fixed bias voltage $V$). From the shape of
$\gamma_{\rm{thr}}$ we can extract several bits of information about
the physical conditions under which the shuttle instability occurs.
First of all, a negative or zero value for $\gamma_{\rm thr}$ means
that no instability can be established. As explained above, some
retardation effect (present for any finite value of $\omega \tau_1$
and $\omega \tau_2$) is necessary in order for the instability to
occur, as the net work of the electrostatic force over one period of
oscillation would otherwise be zero. The smaller the size of the
retardation the more difficult it is to make the system unstable. This
is also confirmed in Fig.~\ref{fig:gamma_thr}, where
$\gamma_{\rm{thr}}$ approaches zero for small values of the
resistances $R_0$ and $R_2$ (implying $\omega \tau_1$, $\omega \tau_2
\rightarrow 0$). On the other hand, it is clear that if the tunneling
of electrons becomes rare due to large junction resistances, the
nanowire is too weakly affected by the electrostatic force to make the
instability occur. This observation is consistent with the behavior of
$\gamma_{\rm{thr}}$ in Fig.~\ref{fig:gamma_thr}, where
$\gamma_{\rm{thr}} \rightarrow 0$ in the limit of large resistances.
Thus, we learn from studying Fig.~\ref{fig:gamma_thr} that the optimal
condition to achieve the shuttle instability is to have the junction
resistances very close to each other, $R_0 \sim R_2$. This implies
$\omega \tau_1 \sim \omega \tau_2 \sim 1$, i.e. quite substantial
retardation effect. Having $Q$-factors between 10 and 1000 and a bias
voltage of the order of \unit[1]{V}, it seems possible to switch on
the instability for a large set of parameters $R_0$, $R_2$. In the
experiment performed by LeRoy \textit{et al.} the junction resistances
are quite different ($R_0 \gg R_2$) \cite{LeRoy2004a} which makes the
electromechanical instability unlikely. For very asymmetric junctions,
the electrostatic potential of the wire at any time is mainly
determined by the strength of the coupling to the lowest resistance.

The physical condition for the occurrence of the shuttle instability
can also be expressed in terms of the bias voltage, $V$, instead of
the dissipation coefficient, $\gamma$. Naturally this is an
experimentally more meaningful quantity, as the amount of dissipation
in the system is difficult to evaluate through only a single
parameter, while the bias voltage can be externally controlled.
\begin{figure}
\begin{center}
\subfigure[]{\label{fig:ampl_CNT}\includegraphics[width=0.35\textwidth]{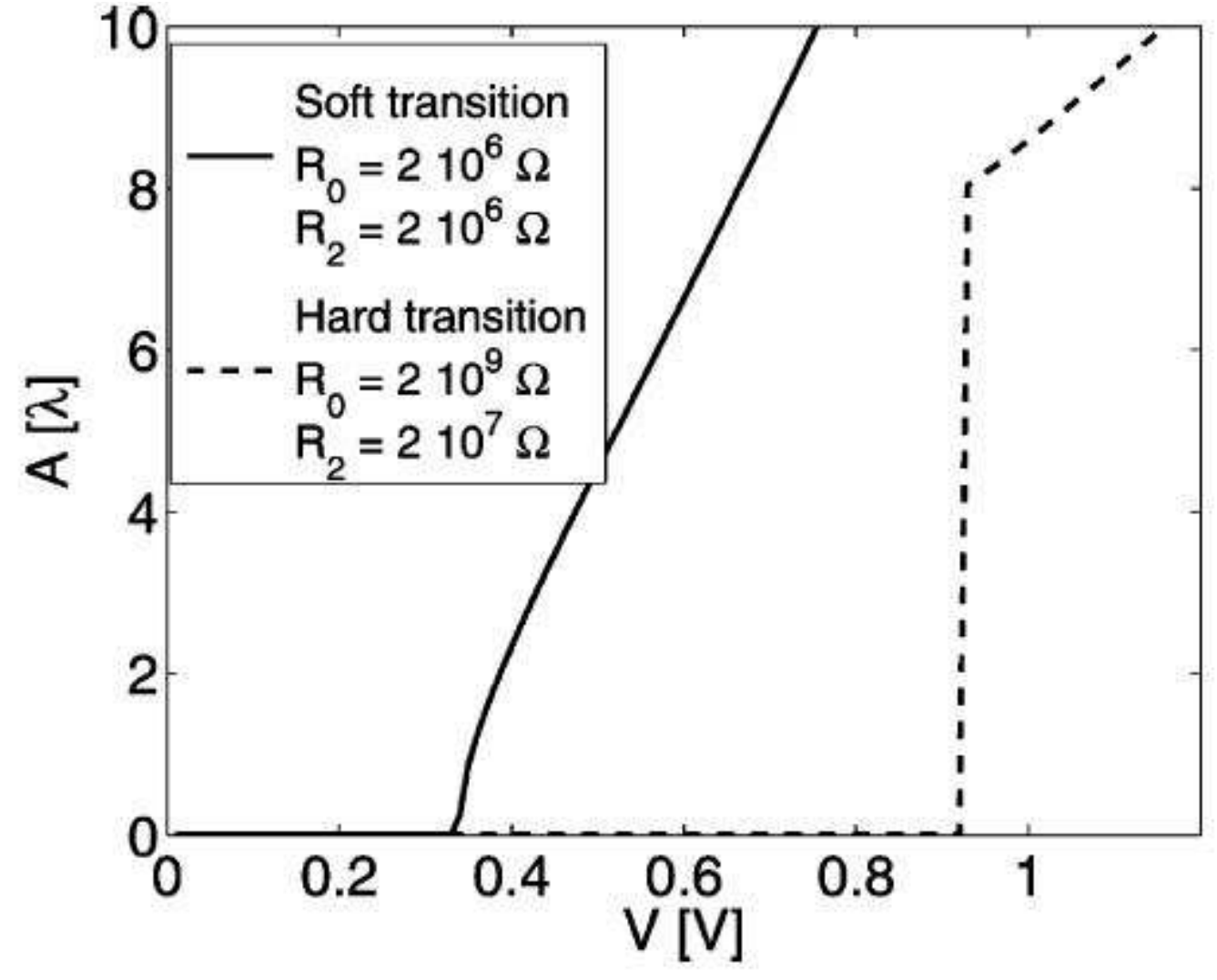}}
\subfigure[]{\label{fig:IV}\includegraphics[width=0.4\textwidth]{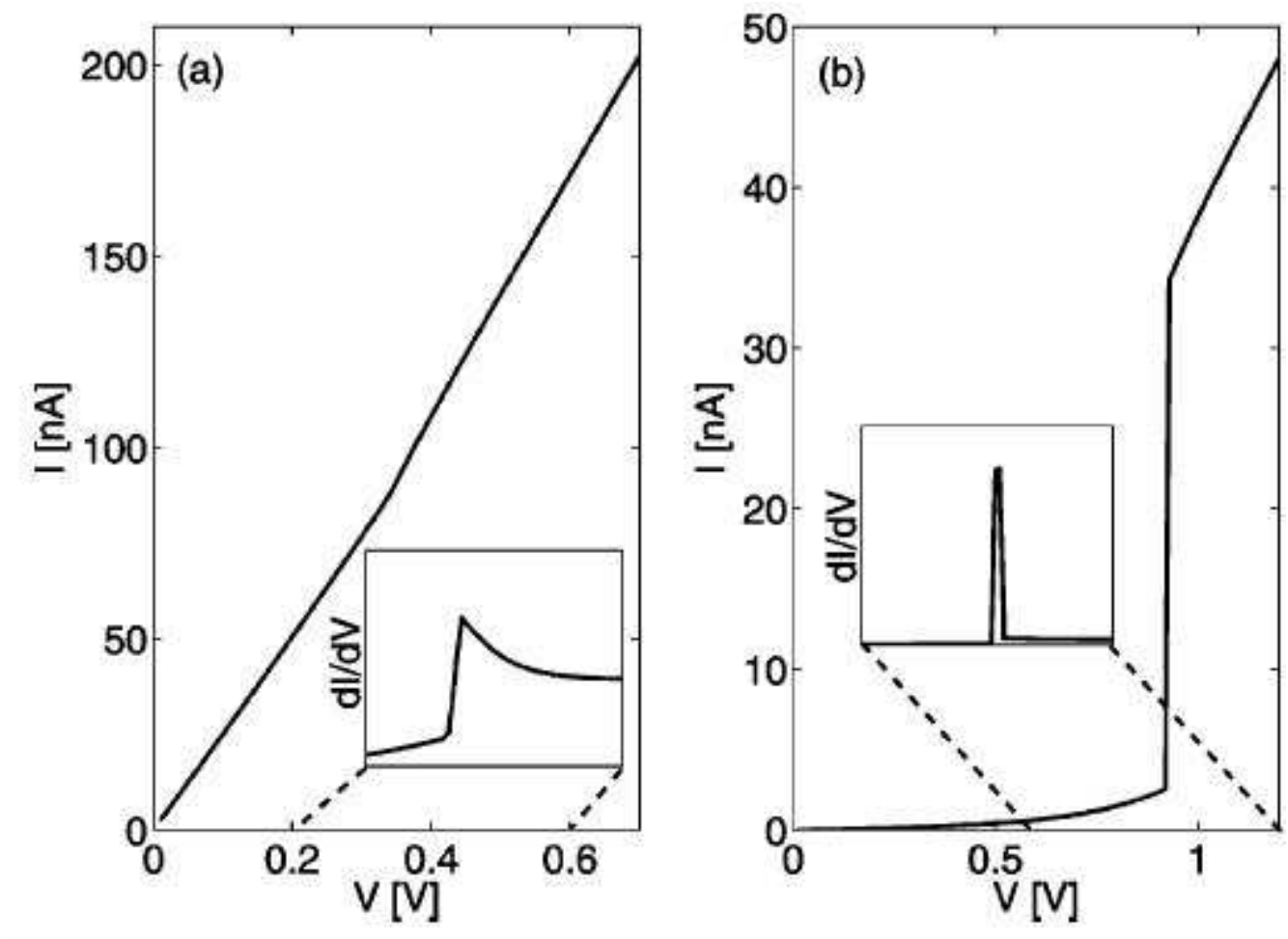}}
\caption{\textbf{(a)} Amplitude of nanowire oscillations as a function
  of the bias voltage. The solid line shows a ``soft'' development of
  the instability, while the dotted line corresponds to a ``hard''
  transition. \textbf{(b)} $I-V$ curves for the same ``soft'' (left)
  and ``hard'' (right) transitions. Insets show corresponding
  $dI/dV-V$ curves. Reprinted with permission from
  [\onlinecite{Jonsson2005}], L. M. Jonsson \textit{et al., Nano
    Lett.}, \textbf{5}, 1165 (2005). $\copyright$ 2005, American
  Chemical Society.}
\end{center}
\end{figure}
The evolution of the amplitude of oscillation as a function of the
applied bias voltage (for a fixed $\gamma$, i.e. fixed quality factor
$Q$) is showed in Fig.~\ref{fig:ampl_CNT}. From this image, the
presence of a threshold voltage above which the instability is
established is clearly visible. The corresponding $I-V$ and
$dI/dV-V$ characteristics have also been calculated according
to the numerical solution of the equations of motion and are shown in
Fig.~\ref{fig:IV}, where the insets refer to the different ways in
which the transition to the new equilibrium state may occur. This
transition depends on the bias voltage and the parameters of the
system, and can be classified as either smooth (``soft'' excitation)
or step-like (``hard'' excitation) \cite{Isacsson1998}. These plots
provide indications of what is expected to be found in an experimental
investigation of the shuttle instability in suspended nano-resonators.

\subsection{\label{subsec:si_multimode} Multimode shuttling of single electrons}

We now generalize the analysis of the system to include the full
multimode description of its mechanical degrees of freedom. The
formal similarity between the standard shuttle model and the
approach used here allows us to include also Coulomb blockade
phenomena, electronic level quantization and quantum description of
the dynamics.

In order to make the system sketched in Fig.~\ref{fig:model}
equivalent to a single electron transistor device we will now consider
a different bias voltage for the two electrodes. The left electrode is
assumed to be at potential $-V/2$, while the right is capacitively
connected such that it forms a gate electrode with potential
$V_{\rm{g}}$. No tunneling is allowed between the nanowire and the
right electrode as in a standard SET device. The analysis of this
model was first presented by Jonsson \textit{et al.}
\cite{Jonsson2007}.

Considering that the nanowire has a finite length, the set of
single-particle states available for electrons tunneling on it is
quantized. We assume that these states are equally spaced in energy by
the amount $\Delta$. Under these conditions, two regimes can be
distinguished according to the size of $\Delta$ compared to the other
characteristic energies of the system (i.e. $eV$ and $k_BT$). There is
a ``continuum'' regime, defined by $eV,k_BT \gg \Delta$, where
quantization is not important, and a ``discrete'' regime, defined by
$eV,k_BT \ll \Delta$, where the electronic states on the tube are so
far apart in energy that only one of them falls within the window set
by the bias voltage. Under the latter conditions, only one state is
involved in the electronic charge transport. As compared to the
previous section, we now describe the tunneling junctions through the
tunneling rates $\Gamma_1$ and $\Gamma_2$, where the subscripts ``1''
and ``2'' refer respectively to the junction between the STM tip and
the wire and to the junction between the wire and the electrodes. The
right tunneling rate, $\Gamma_2$, is assumed to be constant, while the
left rate, $\Gamma_1$, depends on the tube displacement. What differs
the ``continuum'' and ``single-level'' regimes is that the tunneling
rate is voltage-dependent for the former, whereas this is not the case
for the latter. Furthermore, we assume that the temperature is low
enough that ``backward'' tunneling processes (from the right to the
left junction) are not allowed and charging of the tube with more than
one electron is prevented by Coulomb blockade ($k_BT<e^2/2C$, where
$C$ is the capacitance of the nanowire).

In order to study the motion of the system it is convenient to give a
quantitative definition of the electromechanical coupling of the
different mechanical modes. Let us consider the wire initially at rest
in its stationary configuration, with no extra charge on it and let
$L$ be the distance from the STM tip. As soon as the bias voltage is
switched on and an electron tunnels onto the wire the electrostatic
field produced by the STM exerts an attractive force on it. The
nanowire moves from its stationary configuration with the
corresponding displacement, $\delta\bar{u}_0$, being given by the
equation of motion derived from linear elasticity theory,
Eq.~\eqref{eq:elast},
\begin{equation} \label{eq:displacement}
\delta\bar{u}_0 = \frac{e\mathcal{E}}{m}\sum_{j=1} \frac{\varphi_j(z_0)^2}{\omega_j^2}\,.
\end{equation}
Here, the sum is extended to all modes whose shape profile is
antisymmetric with respect to the wire midpoint ($j=1,3,5,
\ldots$). For larger absolute values of $\delta\bar{u}_0$ and smaller
separations between the nanowire and the STM tip the probability of
tunneling increases appreciably. The characteristic length that
provides a natural reference for all the other lengths in the system
is the tunneling length $\lambda$. Therefore, the ratio between the
displacement due to the electrostatic force and the tunneling length
$\delta\bar{u}_0 / \lambda \equiv \varepsilon$, provides an estimate
of how much the motion of the wire is affected by the tunneling of
charge.  The ratio between each term in the sum in
\eqref{eq:displacement} and $\lambda$ can thus be interpreted as the
electromechanical coupling for each mode. The electromechanical
coupling strength $\varepsilon$ is the same parameter that defines the
ratio between the electrostatic force and the elastic force in the
theory of the ordinary shuttle \cite{Gorelik1998}.

Since the shape profiles of the normal modes $\{\varphi_j(z)\}$ are
oscillating functions of the coordinate $z$ whose wavelength decreases
with increasing mode index $j$, the coupling of the higher modes to
the STM tip is weaker than to the lower modes. Therefore, a valid
approximation is to take into account only a limited number $K$ of
flexural modes. In the quantum description developed in
Ref.~[\onlinecite{Jonsson2007}], the electromechanical coupling for
the $j-$th mode is defined as $el_j\mathcal{E}/ \hbar \omega_j
\lambda_j \equiv d_j/\lambda_j$, where $l_j = \sqrt{\hbar /
  m\omega_j}$ and $\lambda_j$ are respectively the zero-point
amplitude and the effective tunneling length for the $j-$th mode.  The
electromechanical coupling can be defined and evaluated also for other
types of mechanical modes, as shown in several works,
Refs.~[\onlinecite{LeRoy2004,Sazonova2004,Izumida2005,Sapmaz2006}].

The dynamics of this system can be described through a generalized
master equation for the two reduced density matrices that describe the
neutral, $\rho_0$, state of the nanowire and the state with one extra
electron, $\rho_1$ (higher charge states are not allowed because of
the Coulomb blockade),
\begin{subequations}
\label{eq:GME}
\begin{align}
\frac{\partial \rho_0}{\partial t}  &= -i[\hat{H}_{osc},\rho_0] + \Gamma_R\rho_1-\frac{1}{2}\{\Gamma_L(\hat{\mathbf{x}},V)\}\rho_1  \nonumber \\*
&-\frac{\gamma}{2}\sum_n(i[\hat{x}_n,\{ \hat{\pi}_n,\rho_0\}] +[\hat{x}_n,[\hat{x}_n,\rho_0]]) \label{eq:rho_0} \\*
\frac{\partial \rho_1}{\partial t} &= -i[\hat{H}_{osc}(\hat{\mathbf{x}}-\mathbf{d}),\rho_1]+ \sqrt{\Gamma_L(\hat{\mathbf{x}},V)}\rho_0\sqrt{\Gamma_L(\hat{\mathbf{x}},V)} \nonumber \\*
-&\Gamma_R\rho_1 -\frac{\gamma}{2}\sum_n(i[\hat{x}_n,\{ \hat{\pi}_n,\rho_1\}] +[\hat{x}_n,[\hat{x}_n,\rho_1]]) \label{eq:rho_1}\,.
\end{align}
\end{subequations}
In Eq.~\eqref{eq:GME}, $\hat{H}_{osc}$ is the quantum mechanical
Hamiltonian of the vibrational modes of the wire that, as in the
classical treatment, form a set of non-interacting harmonic
oscillators,
\begin{equation} \label{eq:H_osc}
\hat{H}_{\textrm{osc}} = \hbar \omega_1 \sum_n \beta_n \left( \frac{\hat{x}_n^2}{2} + \frac{\hat{\pi}_n^2}{2} \right)\,,
\end{equation}
where $\beta_n \equiv \omega_n / \omega_1$ and $\{\hat{\pi}_n\}$ are
the momenta conjugated to the mode amplitudes $\{\hat{x}_n\}$. The
charging effects and the electromechanical coupling between the
nanowire and the STM tip enter into the description of the system
through the term
\begin{equation} \label{eq:H_tube}
\hat{H}_{\textrm{tube}} = \sum_k E(V_g) c^\dagger_k c_k + \frac{U}{2}\hat{n}^2 - \hbar \omega_1 \hat{n} \sum_j \hat{x}_j \varepsilon_j\,,
\end{equation}
with $E(V_g)$ being the single electron energies, $c^\dagger_k$
[$c_k$] are creation [annihilation] operators of single electron
states labeled by the momentum $k$, $U$ is the electrostatic energy of
the nanowire and $\varepsilon_j$ is the electromechanical coupling for
the $j-$th mode. Dissipative effects are here assumed to be relatively
weak and are characterized by a frequency-independent dissipation rate
coefficient $\gamma$ (``Ohmic damping'').

\begin{figure}
\begin{center}
\subfigure[]{\label{fig:ec_weak}\includegraphics[width=0.37\textwidth]{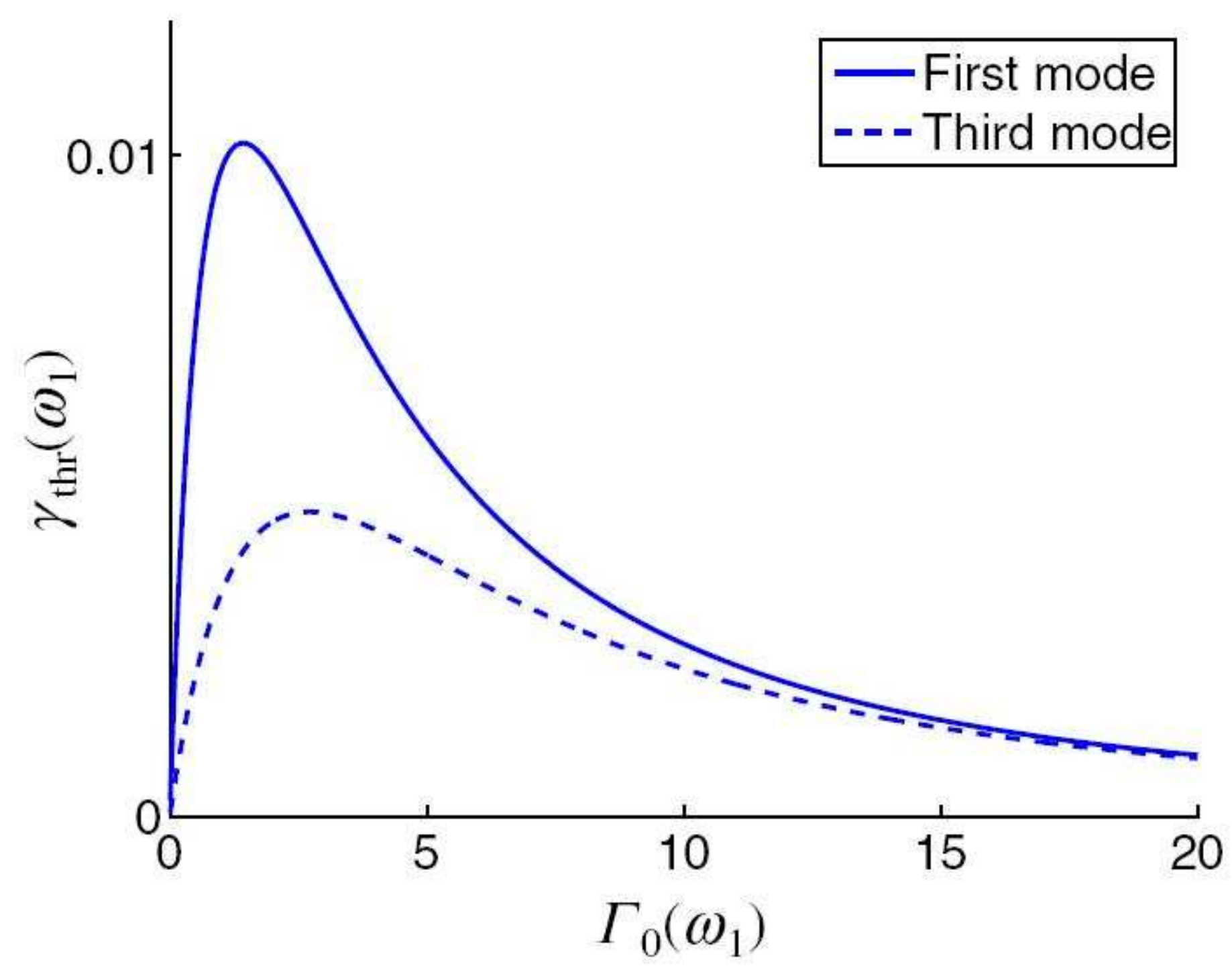}}
\subfigure[]{\label{fig:ec_strong1}\includegraphics[width=0.37\textwidth]{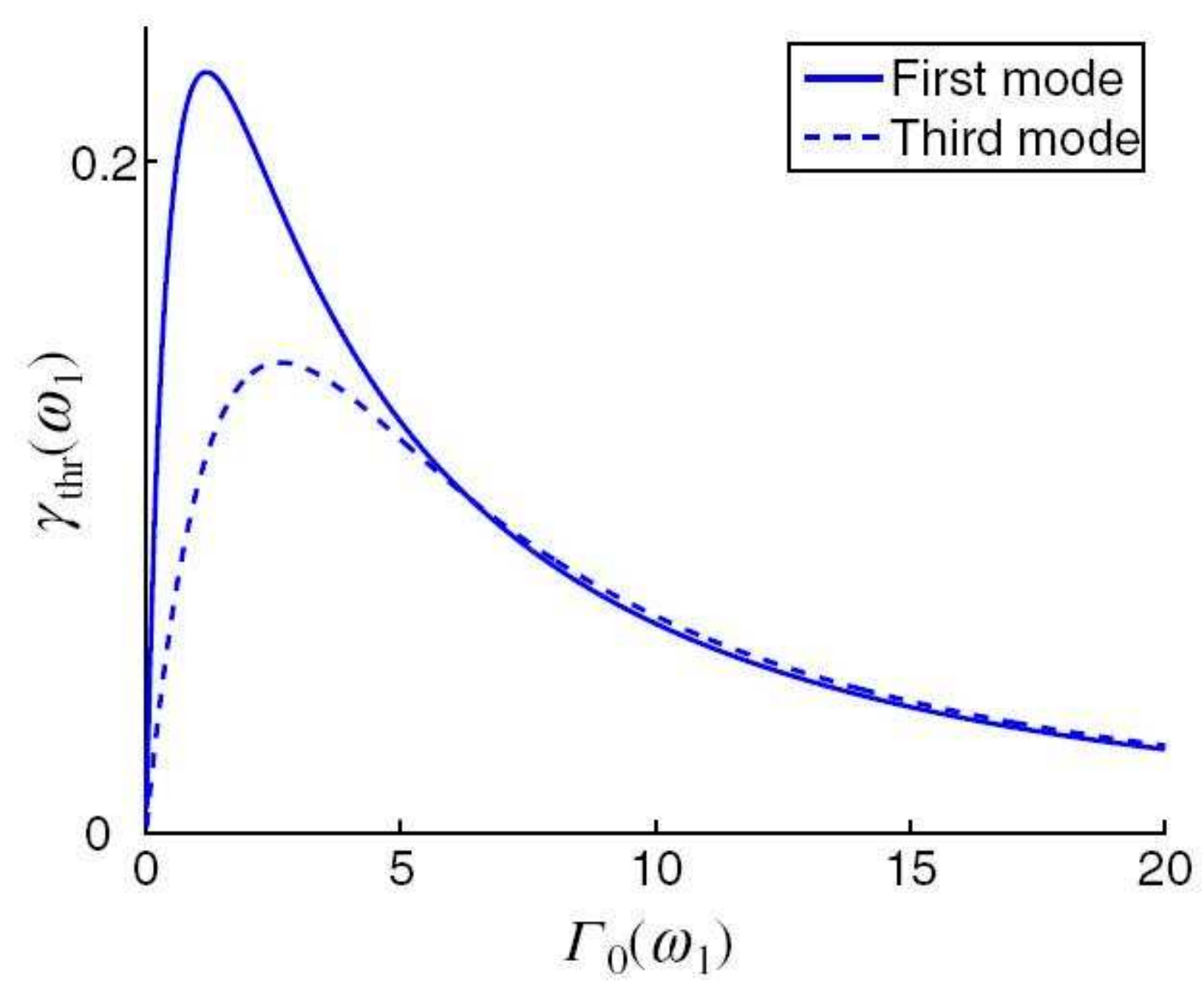}}
\subfigure[]{\label{fig:ec_strong3}\includegraphics[width=0.37\textwidth]{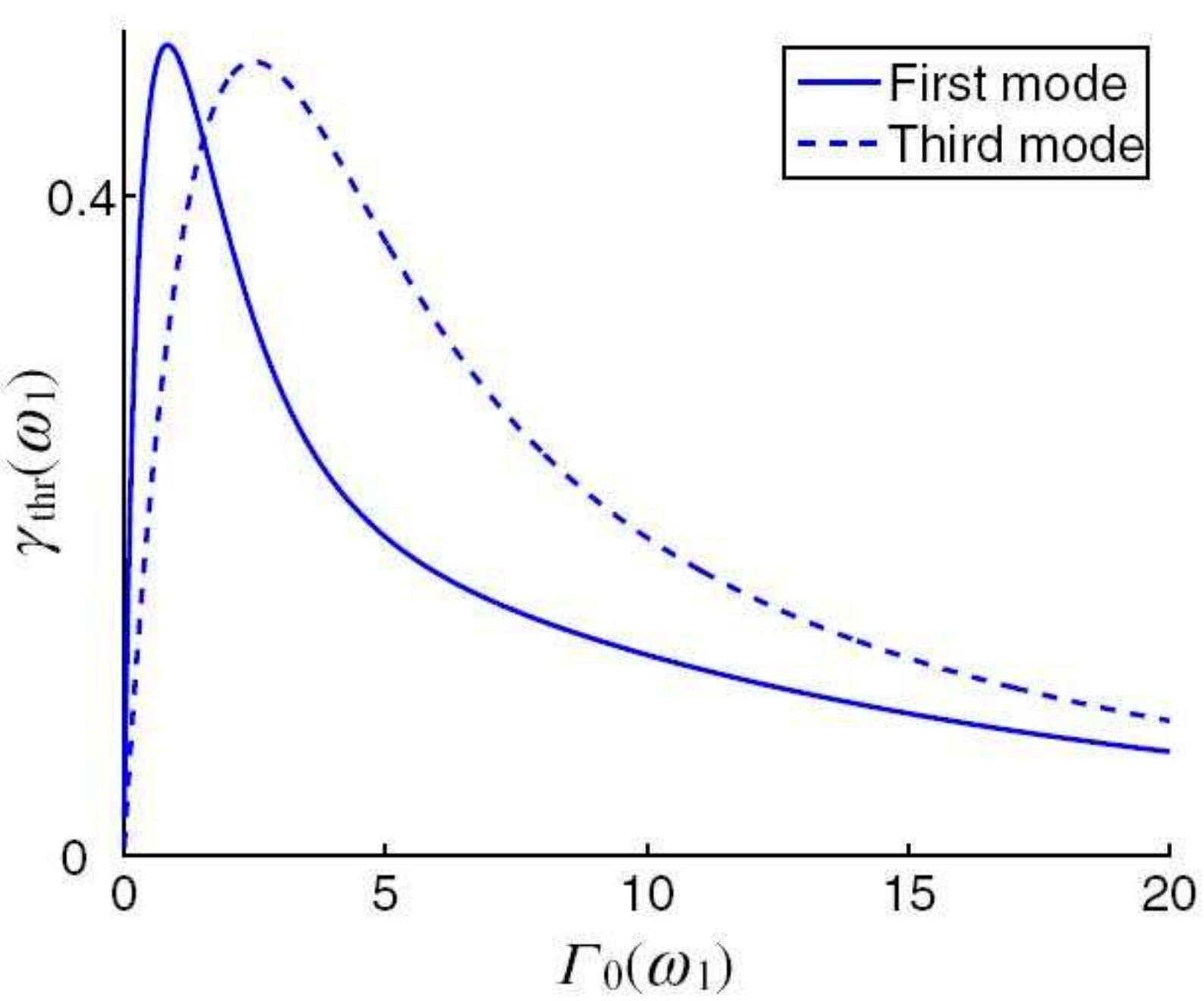}}
\caption{(Color online) Threshold dissipation for the first
  ($\gamma_{\textrm{thr,1}}$) and third ($\gamma_{\textrm{thr,3}}$)
  transverse mode as a function of $\Gamma_0$. Going from top to
  bottom the electromechanical coupling is increased from the weak to
  the strong regime, $d_1/\lambda <1$ \textbf{(a)}, $d_1/\lambda\sim
  1$ \textbf{(b)} and $d_1/\lambda\gg 1$ \textbf{(c)}. When the
  electromechanical coupling is large the interaction between
  different modes is no longer negligible, which introduces a
  qualitatively new feature compared to the weak electromechanical
  regime. As can be seen, some values of $\Gamma_0$ allow for mode
  $n$ to be excited and mode $m$ to be suppressed even though $n>m$,
  i.e. $\gamma_{\textrm{thr,n}}>\gamma_{\textrm{thr,m}}$.  Reprinted
  with permission from [\onlinecite{Jonsson2007}], L. M. Jonsson
  \textit{et al., New J. Phys}, \textbf{9}, 90 (2007). $\copyright$
  (2007), Deutsche Physikalische Gesellschaft.}
\end{center}
\end{figure}
From Eq.~\eqref{eq:GME}, average displacements, momenta and
probabilities can be found from the reduced density matrices and their
respective equation of motion can be derived. These equations can be
linearized around the stationary solution and the behavior of the
solution can be investigated in order to check the onset of the
instability. In the weak electromechanical coupling regime this
analysis can be carried out analytically and an expression for the
threshold dissipation can be found.  The results obtained indicate
that the different modes can be treated independently from each other,
as is shown in Fig.~\ref{fig:ec_weak}. The strong electromechanical
coupling regime must however by analyzed numerically. The consequences
of the transition from the weak to the strong coupling are shown in
Fig.~\ref{fig:ec_strong1} and \ref{fig:ec_strong3} where the threshold
dissipations for mode 1 and 3 are plotted as functions of the
coefficient of the left tunneling rate $\Gamma_0$. These curves
demonstrate the possibility, in the strong coupling regime, to excite
a certain mode $m$ without also making all the modes $n<m$ unstable,
something that is unavoidable in the weak regime. The parameter
$\Gamma_0$ is experimentally controllable as it is determined by the
distance between the STM tip and the nanowire in its stationary
configuration. This distance can be measured and controlled with
sufficient accuracy to make realistic devices.

\begin{figure}
\centering
\includegraphics[width=0.8\linewidth]{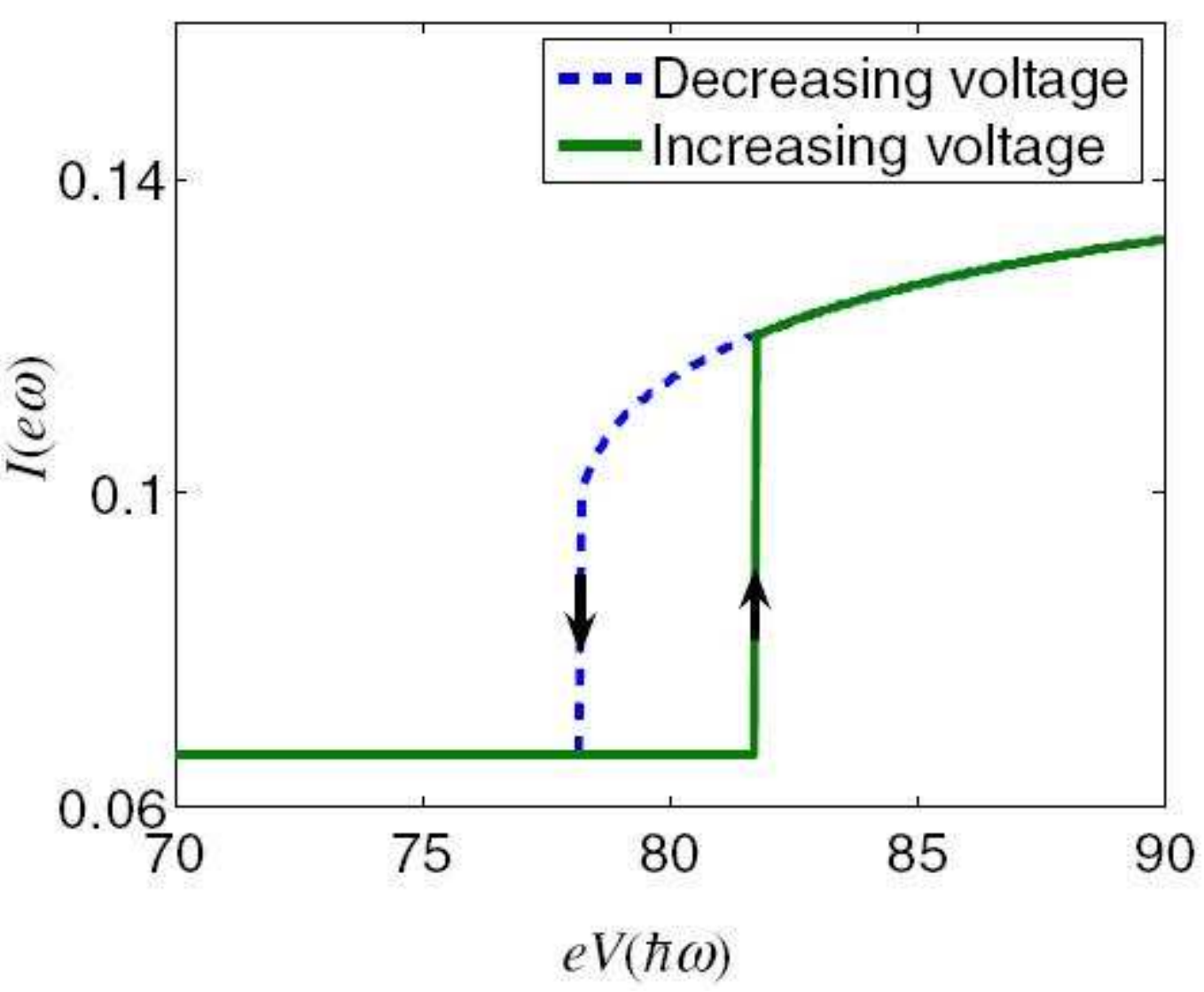}
\caption{(Color online) $I-V$-characteristics for the device of
  Fig.~\ref{fig:model} in the weak coupling regime. For small
  voltages, the current is constant and depends only on the tunneling
  through the double static junction
  $I=e\Gamma_0\Gamma_R/\Gamma_t$. Above the threshold voltage the
  current also depends on the vibration amplitude. Here the
  electromechanical instability occurs through the ``hard''
  transition, characterized by the displayed hysteresis in the
  $I-V$-curve. Reprinted with permission from
  [\onlinecite{Jonsson2007}], L. M. Jonsson \textit{et al., New
    J. Phys}, \textbf{9}, 90 (2007). $\copyright$ (2007), Deutsche
  Physikalische Gesellschaft.}
\label{fig:IV-1_Magnus2}
\end{figure}

A natural issue that needs to be clarified is what happens to the
system once the instability develops. From the theory of the ordinary
shuttle it is known that the amplitude of the oscillations should
increase and then saturate to a value determined by the parameters of
the device, the system reaches a limit cycle.  In order to check the
existence of the limit cycle for the system considered here,
linearization of the equations of motion is not enough and the full
nonlinearity due to the exponential form of the left tunneling rate
must be taken into account. This issue has been investigated in
Ref.~[\onlinecite{Jonsson2007}] for the case of weak coupling and a
single unstable mode. The existence of the limit cycle can be proved
using the Wigner representation of the density operator, following the
approach developed in Ref.~[\onlinecite{Fedorets2004}]. The advantage
of this approach is that it makes the crossover between the tunneling
and shuttling regimes more transparent, as discussed in
Ref.~[\onlinecite{Novotny2003}].  From this analysis it is found that
until the electromechanical instability develops, the system behaves
basically like a series of two tunneling junctions, hence the only
expected contribution to the current comes from tunneling. However,
once the instability sets in and the system reaches its limit cycle
(characterized by steady amplitude oscillations) the current is
drastically modified. In Fig.~\ref{fig:IV-1_Magnus2} the current is
plotted as a function of the electrostatic energy $eV$ for the case of
a single mode and ``hard'' instability transition .
As can be seen in this image the characteristics of the ``hard''
transition are clearly visible in the sudden increase in the amplitude
(and hence current) when the bias voltage exceeds the threshold value.
Another distinctive feature of this process (that is absent for the
case of ``soft'' transition) is \textit{hysteresis}. Sweeping the
voltage down from a value above the threshold, the current does not
correspond to the values found before the onset of the instability.

\subsection{\label{subsec:multistability} Multistability and self-organization of multimode shuttle vibrations}

The consequences of the electromechanical instability on the whole set
of flexural modes (not only on the lowest frequency mode) has been
investigated by Jonsson \textit{et al.} \cite{Jonsson2008}, following
a classical approach. Starting from linear elasticity theory, the
equations of motion for the amplitudes of the transverse normal modes
were found to be (the STM tip is still supposed to be at $l/2$ so only
even modes with respect to the midpoint of the tube are included),
\begin{equation}
\ddot x_n + \gamma \dot x_n + \omega_n^2 x_n = q\mathcal{E} / {m}\,.
\label{eq:x_n}
\end{equation}
Here, the force term that acts on the wire depends on the average
excess charge, $q$, on it and the effective electric field
$\mathcal{E}$ produced by the applied bias voltage between the STM tip
and the electrodes. The dissipation is modeled by a linear viscous
term $-\gamma \dot{x}_n$ and, since the damping coefficient $\gamma$
is assumed to be constant, every mode is damped in the same way.

An important parameter for the system under consideration is the ratio
between the typical mechanical vibration frequency $\omega$ and the
characteristic rate of electron tunneling $\Gamma$. In the following
we assume that tunneling events occur frequently compared to the
period of oscillations, i.e. $\omega \ll \Gamma$. Even though this is
far from being the optimal condition for the instability to occur, as
discussed in Section~\ref{subsec:si_singlemode}, it allows us to write
down a simple kinetic equation for the time evolution of the
probability of having one extra charge on the wire (see below). This
choice will result in more experimentally demanding conditions to
achieve the instability (i.e.  higher $Q$-factors) with respect to
those previously discussed in Section~\ref{subsec:si_singlemode}. A
more general approach will be followed in
Section~\ref{subsec:STM_displacement}.

As was done in Ref.~[\onlinecite{Jonsson2007}], we here assume that
the temperature, the bias voltage and electrostatic charging energy of
the nanowire are such that backward tunneling processes are forbidden
and no more than one extra electron is allowed on the nanowire at any
given time. Let us indicate with $p_1(t)$ and $p_0(t)$ respectively
the probability to find one and zero electron on the nanowire at time
$t$. Then, the average excess charge at time $t$ can be expressed as
$q(t)=ep_1(t)$. The variation of $p_1(t)$ in time is given by the
difference between the probability of tunneling from the STM tip to
the nanowire when the latter is neutral and the probability of
tunneling from the nanowire to the electrodes when there is already
one extra electron on it. We also have that the probabilities for the
allowed charge states must sum up to 1 ($p_1+p_0 = 1$, higher charge
states are prevented by the Coulomb blockade). Exploiting this fact,
we can write down the rate equation for $p(t) \equiv p_1(t)$ in the
form,
\begin{equation} \label{eq:p}
\dot p(t) = -\Gamma (u_0) p(t)  + \Gamma_1 (u_0)\,,
\end{equation}
where $\Gamma (u_0) = \Gamma_1 (u_0) + \Gamma_2$ with $u_0 =
\Sigma_{n} x_n\varphi_n(z_0) \sim \Sigma_{n} x_n$, while $\Gamma_1 (x)
= \Gamma_0 \exp (-x/\lambda)$ is the rate of electron tunneling across
the STM-nanowire junction, so that the typical rate of electron
tunneling is $\Gamma \equiv \Gamma_0 + \Gamma_2$. The solution of
Eq.~\eqref{eq:p} can be written as a series expansion in the parameter
$\omega_1/\Gamma$ which is assumed small.

Equations \eqref{eq:x_n} and \eqref{eq:p} describe the coupled
nanoelectromechanical dynamics of the nanowire-based NEM-SET device.
Since strong bias voltages are not allowed if the excess charge on the
wire must be limited to one electron, it is reasonable to consider the
limit of weak electromechanical coupling. In this regime the
displacement of the nanowire due to a single excess charge is small on
the scale of the tunneling length. Then, as shown in
Ref.~[\onlinecite{Jonsson2007}], the onset of the shuttle instability
occurs independently for each vibrational mode and if
$\omega\ll\Gamma$ the instability is ``soft'' \cite{Isacsson1998}.

In order to analyze the evolution of the system after the instability
occurs, it is not sufficient to linearize the equations of motion
\eqref{eq:x_n} and \eqref{eq:p} around the stationary configuration as
the linear force term in the right hand side of these equations simply
tells if the instability can develop or not.  Therefore, it is
necessary to take into account at least some nonlinear terms. Since
within the regime of weak electromechanical coupling the amplitude at
which the nanowire oscillates is expected to be small compared to the
tunneling length, it is possible to expand the tunneling rate
$\Gamma_1(u_0)$ for small $u_0/\lambda$ and keep contribution up to
the third order. Another approximation that is physically motivated in
the weak coupling regime is the use of the Ansatz $x_{n}(t)= \lambda
A_{n}(t) \sin(\omega_{n}t+\chi_{n}(t))$, where the amplitudes $A_n(t)$
and phases $\chi_n(t)$ are slowly varying functions on the time scale
defined by the period of oscillations.  This difference in
characteristic time scales allows us to replace the Ansatz into
Eq.~\eqref{eq:x_n} and take the time-average over the period of the
first, i.e. longest, mode. In this time interval the amplitudes and
phases for the different modes can be considered constant and be
replaced by their average values. Furthermore, the time variation of
the phases is assumed to be negligible with respect to the
eigenfrequencies of the nanowire, $\dot{\chi}_n \ll \omega_n$, which
means that only the amplitudes $A_n$ are taken into account in the
description of the system. The resulting equation of motion for the
averaged amplitudes are,
\begin{equation} \label{eq:A_n}
\dot {A_n} = \alpha_n A_n \left(\delta_n - A_n^2 - 2 \sum_{m \ne n} A_m^2\right)\,,
\end{equation}
where $\delta_n$ and $\alpha_n$ are defined from the following
combinations of parameters
\begin{subequations} \label{eq:alpha_delta}
\begin{align}
\label{eq:delta}
\delta_n & = 16 \bigg(1  - \frac{4 \gamma \Gamma \lambda}{\omega_n^2 d_n} \bigg) + \mathcal{O}\bigg(\frac{\omega_n^2}{\Gamma^2}\bigg) \\*
\label{eq:alpha}
\alpha_n & = \frac{d_n\omega_n^2}{128 \lambda \Gamma} \left[1+ \mathcal{O} \left(\frac{\omega_n^2}{\Gamma^2}\right) \right]\,.
\end{align}
\end{subequations}
To first order in $\omega_n/\Gamma$ the expressions for $\delta_n$ and
$\alpha_n$ do not depend on the mode index $n$, as the product
$\omega_{n}^2 d_{n}$ is independent on $n$, hence one can replace
$\delta_n \rightarrow \delta$ and $\alpha_n \rightarrow \alpha$.
\begin{figure}
\includegraphics[width=0.8\linewidth]{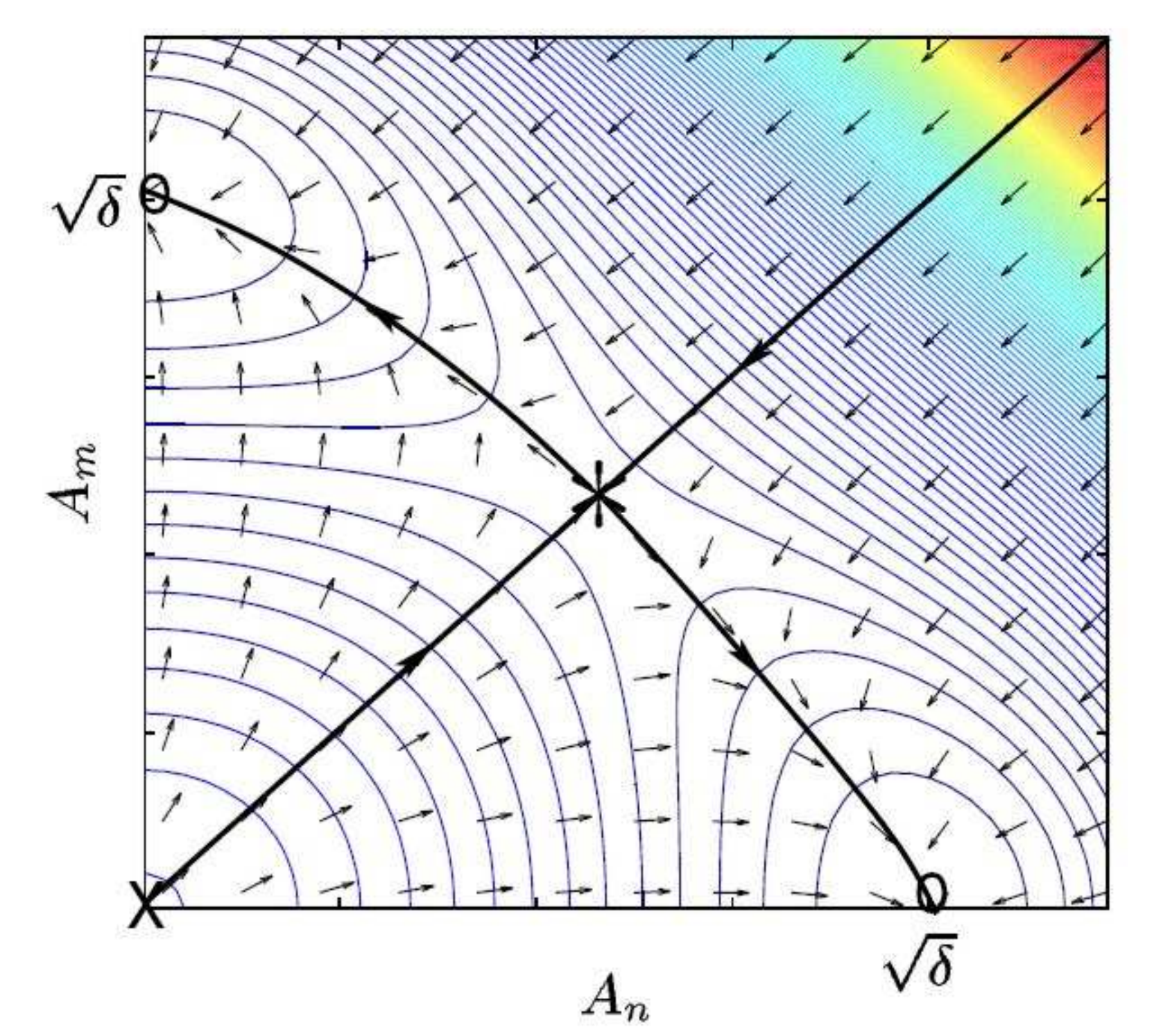}
\caption{(Color online) Stationary points when two modes ($n$ and $m$)
  are unstable. Two attractors, indicated by ({\bf o}),
  corresponding to a finite amplitude of one mode while the other mode
  is suppressed are shown. The stationary point marked with ({\bf x})
  is a repellor and the point indicated by ({\bf $\ast$}) is a saddle
  point. The thick lines are separatrices that trajectories cannot
  cross. The separatrix $A_n = A_m$ ensures that if $A_n(0) > A_m(0)$,
  this inequality hold for all times $t$. Reprinted with permission
  from [\onlinecite{Jonsson2008}], L. M. Jonsson \textit{et al., New
    J. Phys.}, \textbf{9}, 90 (2007). $\copyright$ (2007), Deutsche
  Physikalische Gesellschaft.}
\label{fig:trajectories}
\end{figure}

The behavior of the solution of Eq.~\eqref{eq:A_n} can be explicitly
visualized for the case of two modes $n,m$ (the generalization to more
modes is straightforward). The corresponding stationary points of the
two nonlinear coupled equations can be found analytically and their
stability can be determined through the evaluation of the Jacobian
matrix \cite{Strogatz2001}. As the dynamical behavior of the system
depends on the sign of the parameter $\delta$ we identify, from
Eq.~\eqref{eq:A_n}, that if $\delta<0$ the only stationary point is
the origin, corresponding to the absence of any oscillation. For this
set of parameters the nanowire is at rest in some static configuration
determined only by the constant tunneling rates $\Gamma_0$ and
$\Gamma_2$. On the other hand, if $\delta>0$ the origin becomes
unstable and three more stationary points appear: a saddle point at
($\delta/3, \delta/3$) and two stable points at ($\sqrt{\delta},0$)
and ($0,\sqrt{\delta}$). These two new stable points represent
oscillating states with finite amplitude $\sqrt{\delta}$ and frequency
$\omega_n$ or $\omega_m$ respectively. Which of these two will be
reached by the system depends on the initial conditions as shown in
Fig.~\ref{fig:trajectories}.

The conditions $\delta>0$ defines the onset of the shuttle
instability. From the solution to Eq.~\eqref{eq:A_n} with $\delta=0$
it is possible to find the expression for the threshold electric field
above which the instability starts to develop, $\mathcal{E}_c \equiv
4\Gamma \lambda \gamma m/e$. Further analysis of Eq.~\eqref{eq:A_n}
indicates that once the instability for a certain number of
vibrational modes is established, the system evolves in such a way
that only one of the unstable modes reaches the new stationary state,
characterized by steady amplitude oscillations, i.e. the \textit{limit
  cycle}. It is also found that the selection of the surviving mode is
determined by the initial conditions as the mode which initially has
the largest displacement from the origin (that is from the static
equilibrium state) maintains its separation from the other modes and
evolves into the limit cycle. This can be understood as trajectories
in the amplitude space cannot cross, a result that can be analytically
generalized to an arbitrary number of modes by studying the asymptotic
behavior of the solutions of Eq.~\eqref{eq:A_n} \cite{Jonsson2008}.

It is worth to remark that the symmetry between the modes that
characterizes Fig.~\ref{fig:trajectories} is actually broken if higher
order (in $\omega_n/\Gamma$ and $\omega_m/\Gamma$) corrections to
$\delta$ become relevant or if the dissipation affects each mode in a
different way. Then each mode will have its own $\delta_n$ and, in
general, the selection of the surviving mode may not only depend on
the size of the initial displacements from the static equilibrium
point.

\begin{figure}
\begin{center}
\subfigure[]{\label{fig:comparison_small}\includegraphics[width=0.48\textwidth]{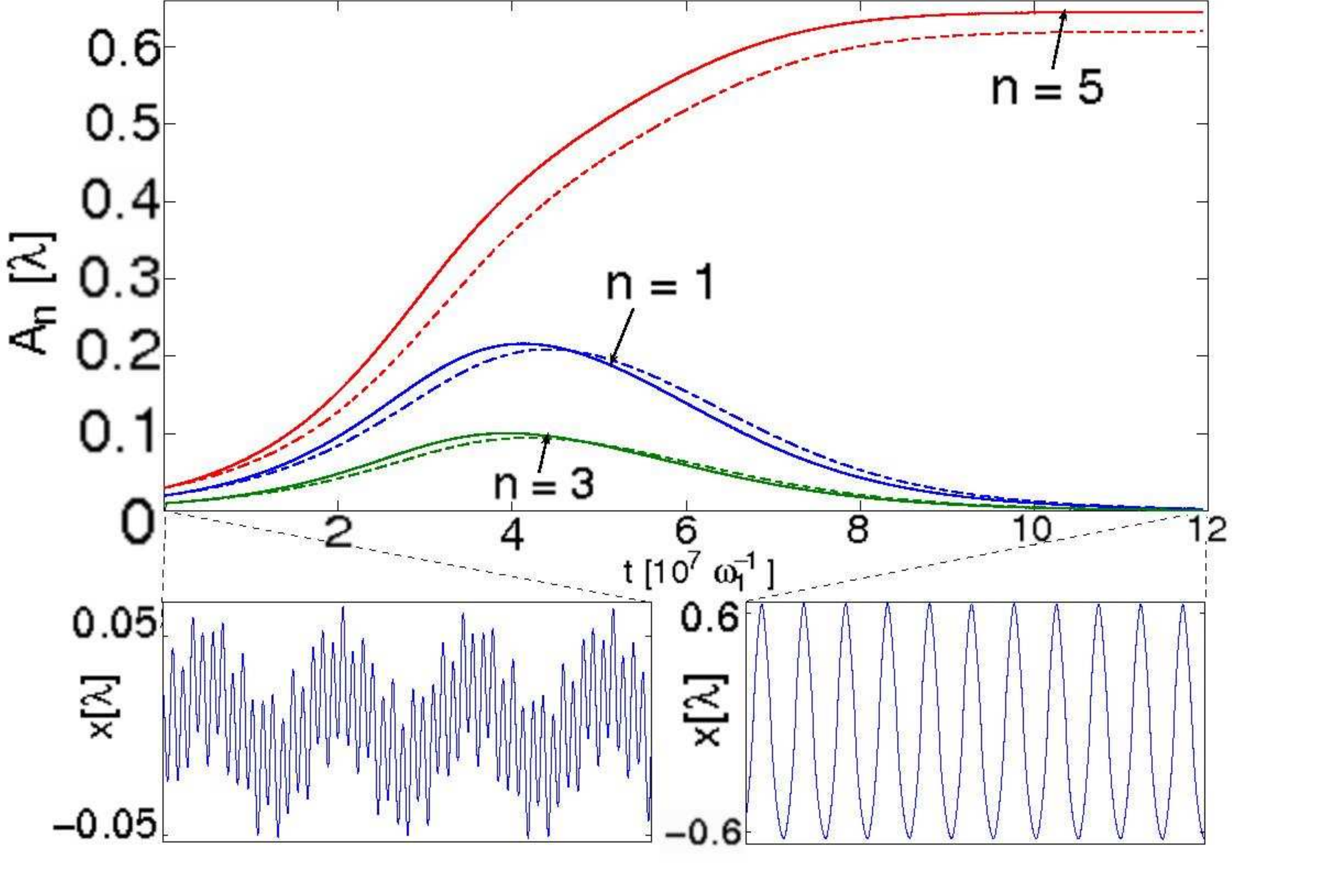}}
\subfigure[]{\label{fig:comparison_large}\includegraphics[width=0.48\textwidth]{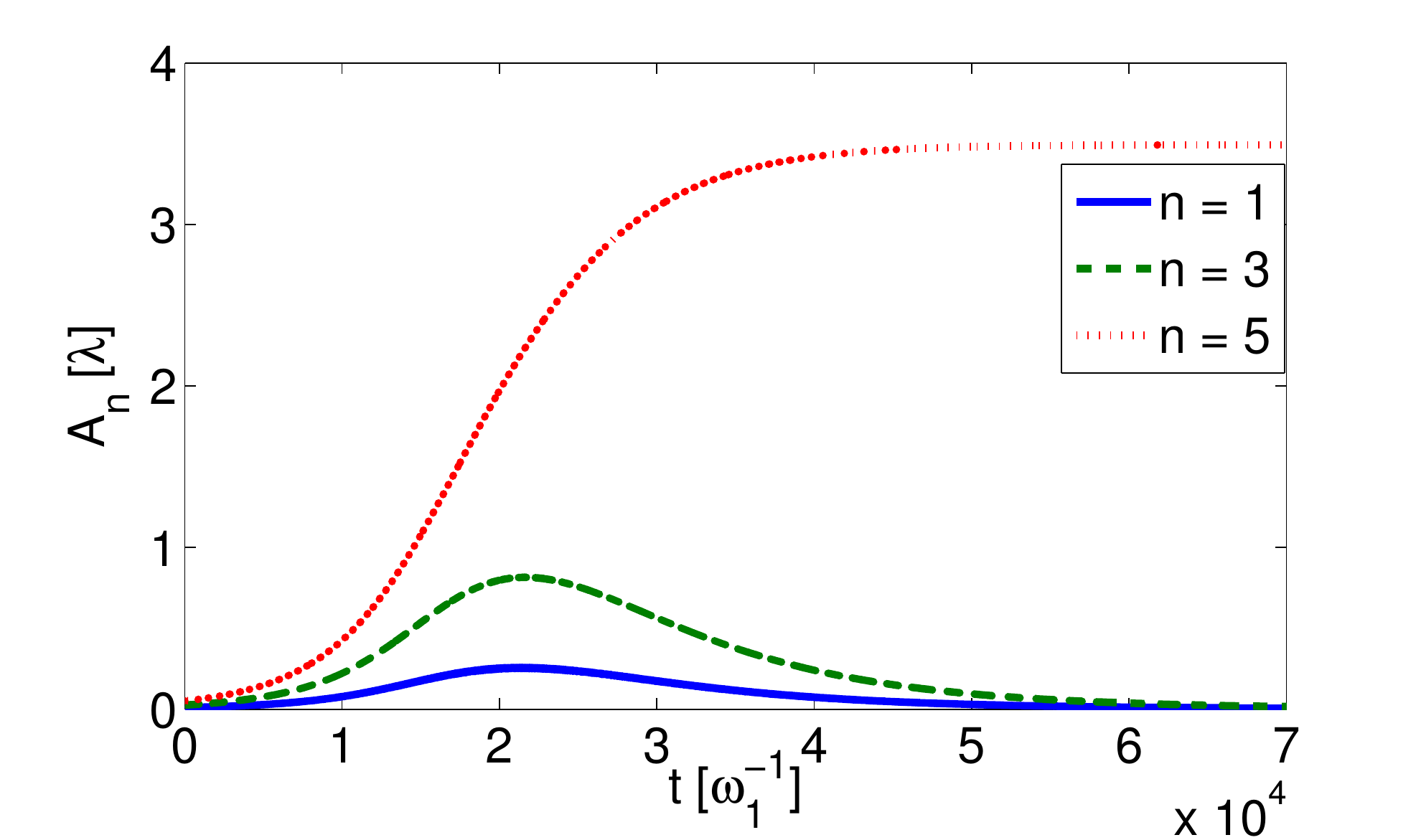}}
\caption{(Color online) Numerical solution of \eqref{eq:x_n} and
  \eqref{eq:p} for the nanotube vibration amplitude as a function of
  time when three modes ($n=1,3,5$) are unstable ($\delta >
  0$). \textbf{(a)} Weak electromechanical regime. Comparison with the
  approximate result, Eq.~\eqref{eq:A_n}, is shown as dashed curves.
  The lower left panel shows the quasi-periodic oscillation of the
  nanotube center position just after the onset of the instability,
  while the lower right panel shows the regular vibrations that appear
  after all but the $n$=5 mode amplitudes have been suppressed (see
  text). \textbf{(b)} Strong coupling regime. The large amplitudes
  make an approximate analysis based on Eq.~\eqref{eq:A_n} invalid,
  but the phenomenon of selective excitation persists. Reprinted with
  permission from [\onlinecite{Jonsson2008}], L. M. Jonsson \textit{et
    al., New J. Phys.}, \textbf{9}, 90 (2007). $\copyright$ (2007),
  Deutsche Physikalische Gesellschaft.}
\end{center}
\end{figure}

In order to check the results obtained by the analysis of
Eq.~\eqref{eq:A_n}, one can also compare them to the numerical
solution of the ``full'' equations of motion, Eq.~\eqref{eq:x_n}. The
comparison for a given choice of initial conditions is shown in
Fig.~\ref{fig:comparison_small}.
Since the frequencies of the different modes are not commensurable,
i.e. they are not integer multiples of the fundamental mode, the
initial motion of the nanowire is not characterized by a sharply
defined periodicity as can be seen in the lower left panel of
Fig.~\ref{fig:comparison_small}. However, when the system reaches the
final stationary state and only the mode $n=5$ (which initially had
the largest deviation away from static equilibrium) has a non-zero
amplitude, its oscillations are clearly periodic with frequency
$\omega_5$. Similar figures with modes $n=1$ or $3$ having finite
amplitudes can be obtained by changing the initial conditions.

Note that the numerical analysis of Eq.~\eqref{eq:x_n} is not
necessarily limited by the requirement that the oscillation amplitude
be kept small with respect to the tunneling length.
Fig.~\ref{fig:comparison_large}, for instance, suggests that the
selective evolution promoted by the simultaneous instability of many
vibrational modes can characterize also the ``large amplitudes''
regime.

\subsection{\label{subsec:STM_displacement}  Geometrical scanning of the flexural modes by STM tip displacement}

The theoretical study described in the previous section suggests that
the development of the shuttle instability in an extended object, such
as a suspended nanowire, induces a selective evolution of its many
mechanical degrees of freedom. It should be stressed however, that the
results presented in Ref.~[\onlinecite{Jonsson2008}] provide more a
theoretical demonstration of this selectivity rather than an
experimental procedure to control it. The prime reason for this is
that the analysis involves the separate choice of initial conditions
for each vibrational mode, an operation that cannot be performed
experimentally. However, the device considered can still be used to
accurately detect the instability of the survivor mode, as the
shuttling current is proportional to the mechanical frequency once the
system reaches the stable limit cycle \cite{Gorelik1998}.  Thus, by
measuring the current, we are able to tell which vibrational mode has
become unstable. 

Besides this, the question of finding a practical procedure to select
single vibrational modes through the shuttle instability naturally
arises. A possible way to do this was suggested in
Ref.~[\onlinecite{Santandrea}], by generalizing some of the features
of the system considered in
Refs.~[\onlinecite{Jonsson2005,Jonsson2007,Jonsson2008}]. In this work
the tip of the STM is specified in a generic position $z_0$ along the
nanowire axis and is no longer fixed above the wire's midpoint.  Under
these circumstances the coupling $\varepsilon$ is affected by the
shape profiles of the eigenmodes at point $z_0$ (see
Eq.~\eqref{eq:displacement}).

Ref.~[\onlinecite{Santandrea}] also considers the effects of
dissipation in the studied device to greater detail than what has been
presented in the previous sections.  Damping of oscillations in solids
can be caused by many different microscopic mechanisms and in general
it is impossible to formulate a theory that is able to describe all of
them from first principles with arbitrary accuracy. In most cases it
is even difficult to accurately determine which is the dominant
mechanism for dissipation. For example, certain dissipative processes
that can be neglected in bulk materials can be enhanced in
nano-devices due to the increased surface-to-volume ratio
\cite{Carr1999}. To account for these considerations
Ref.~[\onlinecite{Santandrea}] analyzes not only the usual ``viscous''
damping term $-\gamma\dot{u}$ but also considers the Zener theory of
dissipation for the standard linear solid \cite{Nowick1972}.

The Zener model provides a simple approach that goes a step beyond the
ordinary elasticity theory. According to Hooke's law, the stress and
strain fields, i.e. the macroscopic variables that characterize the
mechanical state of a solid, are connected by a simple proportionality
relation, such that any change in the former reflects itself
\textit{instantaneously} as a change in the latter and vice versa. In
the Zener theory, it is instead assumed that these quantities need a
finite relaxation time, $\tau_{\sigma}$ and $\tau_{\varepsilon}$
respectively, to reach their equilibrium value. This is a consequence
of the internal processes that dissipate energy during any
modification of the mechanical state of the solid. For the case of
periodic variations of the stress and strain with frequency $\omega$,
the following nonlinear frequency-dependent expression for the
$Q$-factor can be derived from the Zener theory,
\begin{equation} \label{def:Q-Zener}
Q_Z(\omega) = \frac{1}{\Delta} \frac{1+(\omega\bar{\tau})^2}{\omega \bar{\tau}}\,.
\end{equation}
Here, $\bar{\tau}=\sqrt{\tau_\varepsilon \tau_\sigma}$ and $\Delta
\equiv E_U-E_R$ is the difference between the ``unrelaxed'' and
``relaxed'' values of the Young's modulus \cite{Nowick1972}, in effect a
measure of the degree of non-elasticity of the body.

In Ref.~[\onlinecite{Santandrea}] a classical description of the
system depicted in Fig.~\ref{fig:model} was formulated where the
tunneling process that charges the nanowire and makes it sensitive to
the electrostatic force is a stochastic process. As such, the normal
mode amplitudes $\{ x_n \}$ and conjugated momenta $\{ \pi_n \}$ are
also stochastic variables and we define probability densities for them
when the nanowire is charged $P_1(x_1, \pi_1, x_2, \pi_2, \ldots,t)$
and when it is neutral $P_0(x_1, \pi_1, x_2, \pi_2, \ldots,t)$. The
evolution of these objects is determined by a generalized Boltzmann
equation where the ``collisional integral'' is replaced by tunneling
terms following the approach used in Ref.~[\onlinecite{Armour2004}],
\begin{align}
\frac{\partial P_1}{\partial t} & + \sum_i \left[ \frac{\partial (\dot{x_j}P_1)}{x_i}+\frac{\partial (\dot{\pi_j}P_1)}{\pi_j} \right]  \nonumber \\*
&=  \Gamma_1(\mathbf{x},z_0)P_0-\Gamma_2P_1 \label{eq:P1} \\*
\frac{\partial P_0}{\partial t} & + \sum_j \left[ \frac{\partial (\dot{x_j}P_0)}{x_j}+\frac{\partial (\dot{\pi_j}P_0)}{\pi_j} \right] \nonumber \\*
&= -\bigg(\Gamma_1(\mathbf{x},z_0)P_0-\Gamma_2P_1\bigg)\,.
\label{eq:P0}
\end{align}
The behavior of the system is described by the dynamical variables
obtained by averaging the mode amplitudes and conjugated momenta over
$P_0$ and $P_1$: $\langle \ldots \rangle_+ \equiv \langle \ldots \rangle_0 +
\langle \ldots \rangle_1$.

\begin{figure}
\begin{center}
\subfigure[]{\label{fig:instability_viscous}\includegraphics[width=0.48\textwidth]{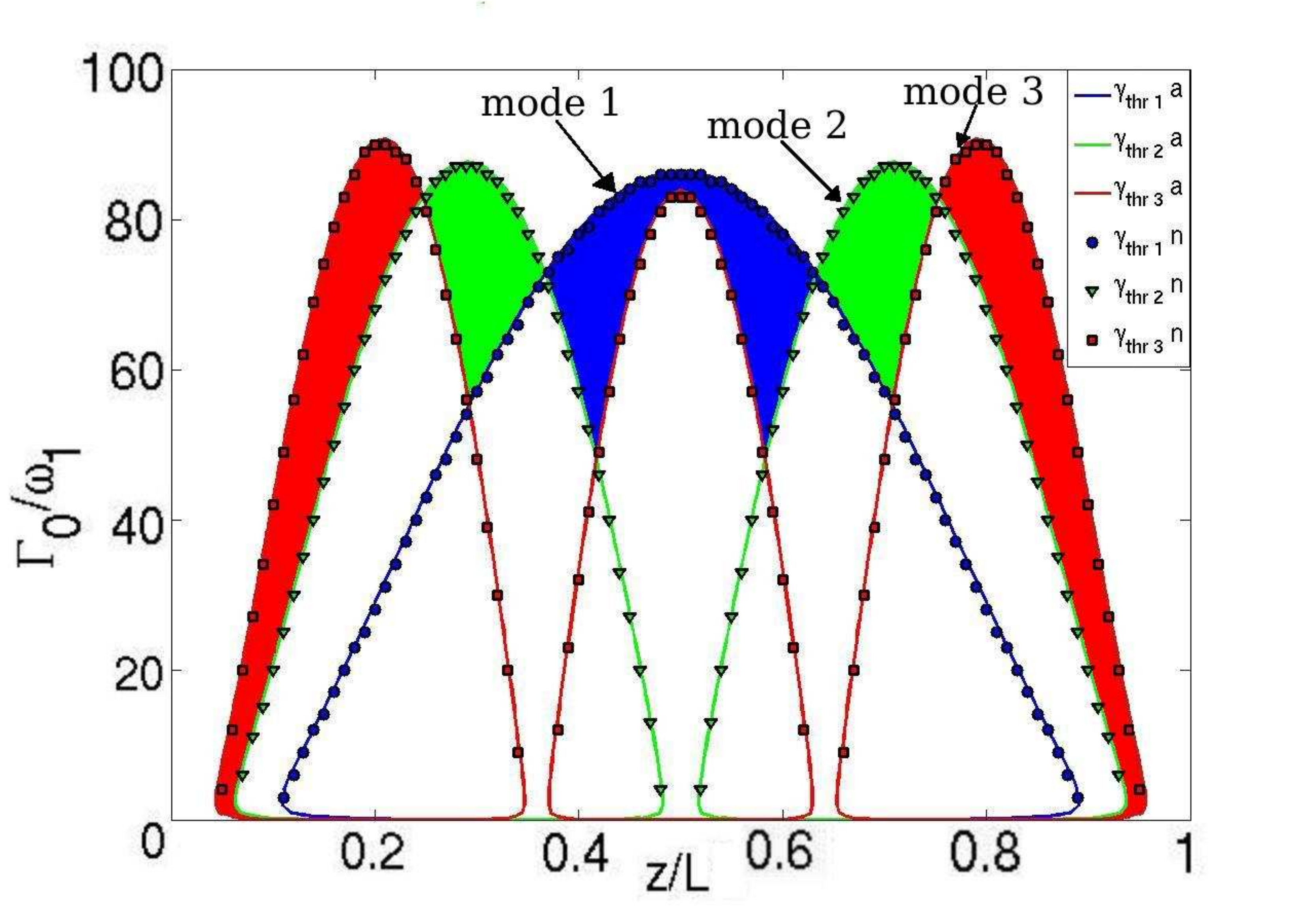}}
\subfigure[]{\label{fig:instability_Zener}\includegraphics[width=0.48\textwidth]{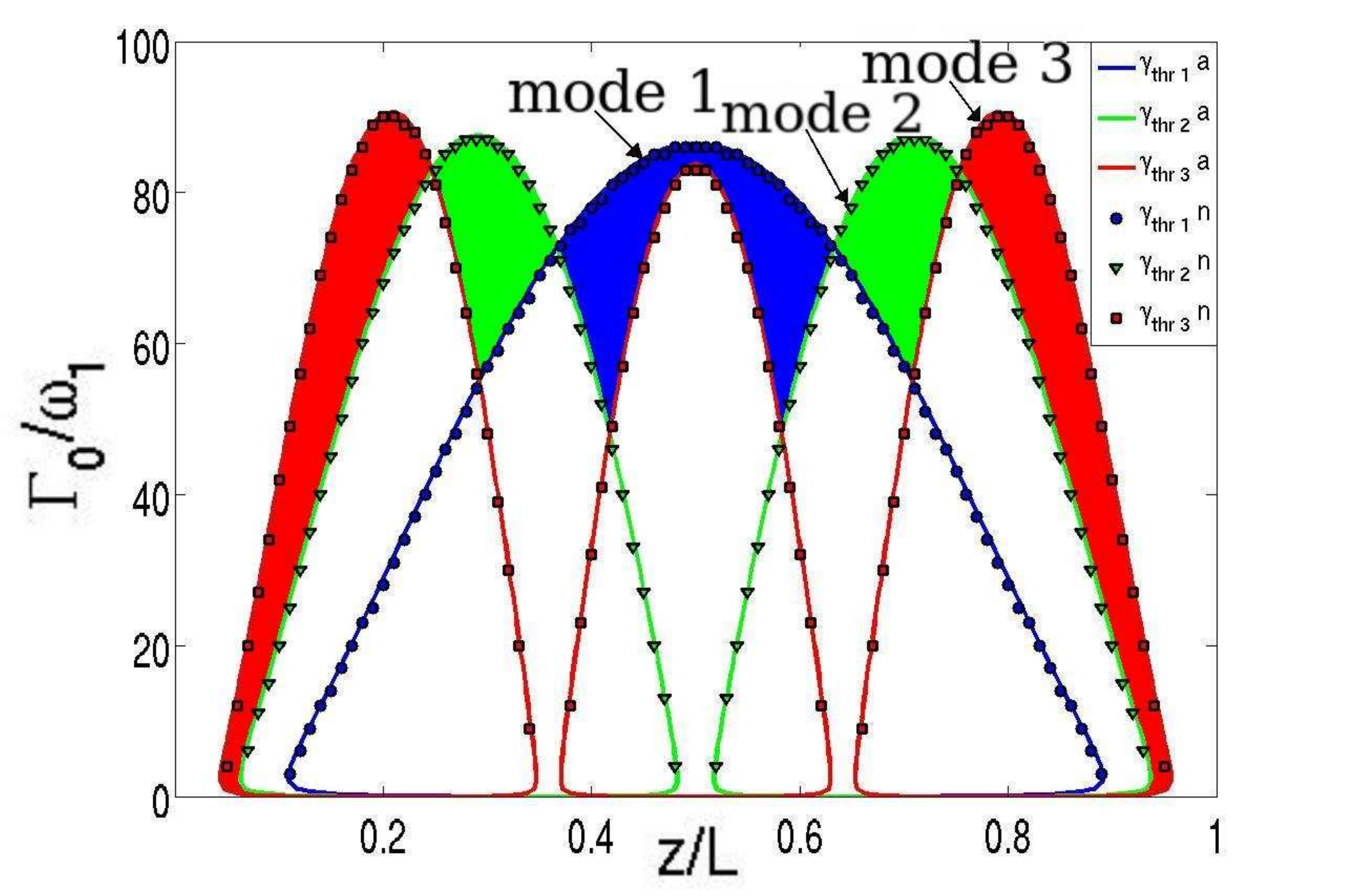}}
\caption{(Color online) \textbf{(a)} Regions of instability for
  dissipation modeled by a viscous term $-\gamma \dot{u}$.
  \textbf{(b)} Regions of instability for internal dissipation given
  by the Zener model. The solid line and the dots that define the
  threshold dissipation curves are obtained respectively from the
  analytic and the numerical solution of the linearized equations of
  motion. In both plots the filled areas define the values of $z$ and
  $\Gamma_0$ for which only a single mode is unstable. Reprinted with
  permission from [\onlinecite{Santandrea}], F. Santandrea}
\end{center}
\end{figure}

The possibility of having a shuttle instability in the weak
electromechanical coupling regime is investigated by linearizing the
equations of motion for the averaged dynamical variables around the
static equilibrium position where the fundamental solutions have the
form $\langle x_j\rangle_+\sim e^{i\omega_j+\delta_j}$. The condition
for instability of the static configuration of the $j$-th mode can
then be expressed as, $\mathfrak{Re}[{\delta_j}]>0$. For a given
quality factor $Q_j$ (that is, for a given amount of dissipation), the
exponent $\delta_j$ for each mode can be plotted as a function of the
two experimentally accessible parameters: the tunneling rate
$\Gamma_0$ between the STM tip and the nanowire in the static
configuration and the position of the STM tip along the wire axis.
The result of this analysis is plotted in
Fig.~\ref{fig:instability_viscous} for the case of viscous dissipation
($Q_j \propto \omega_j$) and in Fig.~\ref{fig:instability_Zener} for
the case of internal dissipation.
The two plots, Fig.~\ref{fig:instability_viscous} and
\ref{fig:instability_Zener}, do not show any qualitative
difference. In both cases one can distinguish sets of parameters
($\Gamma_0$, $z$) for which only one mode is unstable and sets for
which two or more modes are unstable. This fact suggests a general way
to express the condition for shuttle instability for other types of
mechanical degrees of freedom and dissipative processes,
$\mathfrak{Re}[{\delta_j}]>0$, where the real part of $\delta_j$ is
given by,
\begin{equation} \label{eq:instability_general}
\mathfrak{\delta_j} = -\frac{\omega_j}{2Q_j} + \frac{\Gamma_0\Gamma_2}{2\Gamma_t} \frac{e\mathcal{E}}{m\lambda\Gamma_t} \frac{\varphi_j^2(z_0)}{\omega_j^2+\Gamma_t^2}\,.
\end{equation}

\section{\label{sec:NEM} Magnetic field induced NEM coupling}

In the previous sections we have shown that NEM-SET shuttle structures
can be used to significantly alter the electronic and mechanical
characteristics of the discussed system. In what follows, we will
consider further the case of suspended nanowire structures and show
that these can also be made to mechanically oscillate through the
introduction of external magnetic fields. Due to the effective induced
electron-vibron coupling of these devices novel physical phenomena are
predicted as presented below.\\

Although suspended nanowires can be viewed as a particular realization
of a NEM-SET shuttle structure, specific nanoelectromechanical
operation is expected as a result of strong elongation of the movable
part of these devices. This comes about as the very large aspect
ratios in these systems affect both the mechanics of the flexural
vibrations and the electrodynamics of the electronic current
flow. Elongation of a suspended wire, or alternatively diminishing its
cross-section, will make the wire more flexible and hence more
sensitive to external mechanical perturbations. This will in turn
make quantum effects more pronounced due to the larger amplitude of
the wire's zero point flexural vibrations. Utilizing this, new types
of NEM coupling can be achieved through the induced electron-vibron
interaction caused by, e.g., externally applied magnetic fields. This
coupling specifies the Lorentz force acting on the wire, and for case
of mechanically oscillating wires, also the electromotive force on the
electrons which counteracts the motion causing the vibrations. Since
the flexibility of the wire increases with its aspect ratio, and the
Lorenz force is directly proportional to the length of the wire,
significant inductive NEM coupling can be achieved as a result of
strong electrical current concentrations in wires of nanosized
cross-section.

In comparison to the nanoelectromechanics considered in the previous
parts of this paper --- where the tuning of the NEM performance was
achieved through the coupling of an external electric field to the
local charge concentration --- we here show that similar results can
be achieved if external magnetic fields coupled to the electronic
current are instead used. The effects of this type of NEM coupling on
a suspended nanowire of length $L$ and cross-section $S$, carrying a
current $I$ in a transverse magnetic field $H$ is described through
the induced deflection of the wire, $x_d(H)$. From standard elasticity
theory one finds that $x_d(H)\propto Hj/S\alpha(L)$ where $\alpha(L)=
E/L^4$ with $j$ the current density and $E$ the Young's modulus of the
wire. For suspended nanowires of length $L\sim$~\unit[1]{$\mu$m} we
estimate that such deflections can be as large as
$X_d(H)\sim$~\unit[0.1-1]{nm}, which can crucially affect the
electronic tunneling through such mesoscopic NEM structures.  Quantum
coherence in the flexural vibrations of the wire may, as such, affect
the electronic transport in non-trivial ways if the area $Lx_0$ ($x_0$
is the zero point amplitude of oscillation of the wire) available for
penetration of the external magnetic field is comparable to the flux
quantum. For the realistic case of a vibrating carbon nanotube, such
quantum coherent nanoelectromechanics can be achieved for magnetic
fields of the order of a few tens of tesla, corresponding to present
state of the art experimental achievements. In the following sections
two examples of the above discussed inductive nanoelectromechanics
will be presented. The new physics of entanglement of resonantly
tunneling electrons with nanowire quantum vibrations will be presented
in the Section~\ref{sec:Aharonov}, whereas the possibility to pump
nanovibrations by a supercurrent flowing through a suspended nanowire
is discussed in Section~\ref{sec:Super}.

\subsection{\label{sec:Aharonov} Quantum mechanically induced electronic Aharonov-Bohm interference}

\begin{figure}
\includegraphics[width=0.4\textwidth]{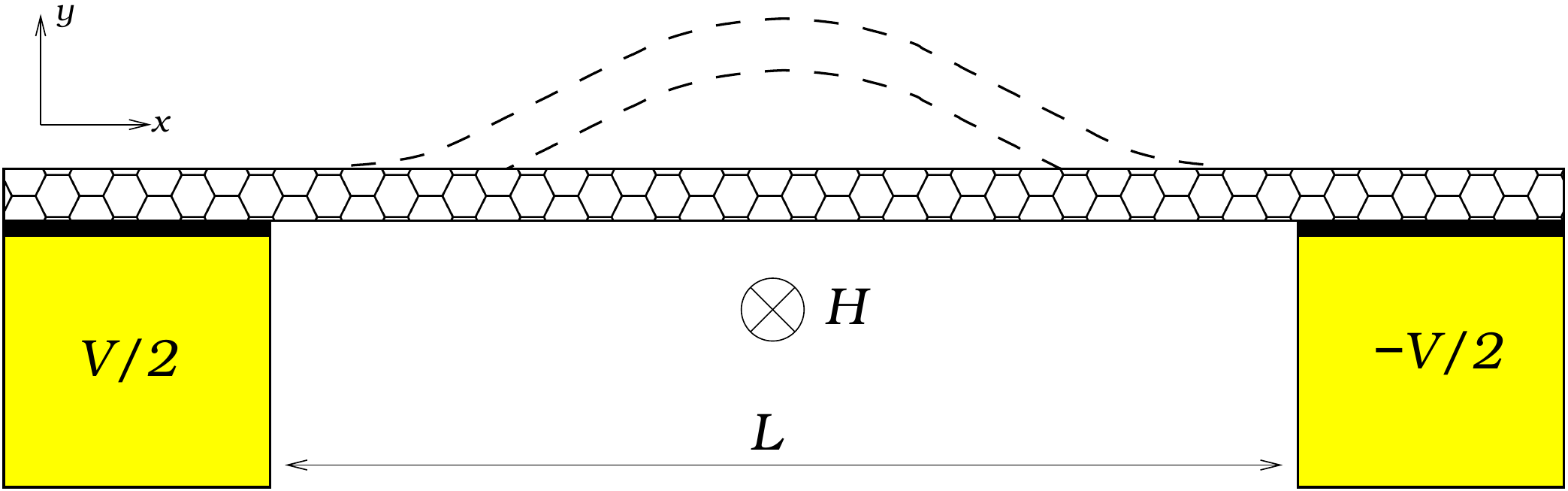}
\caption{Schematic diagram of the system considered in
  Section~\ref{sec:Aharonov}. A transverse magnetic field, $H$, is
  applied to a suspended 1-dimensional model nanowire of length
  $L$. If the wire is biased by a voltage $V$, it carries a current
  and the wire oscillates in response to the induced Lorentz
  force. Quantum fluctuations in the wire's bending modes make the
  electrons propagate along an effectively 2-dimensional wire. The
  magnetic field-induced reduction of electron propagation in the
  elastic channel can be interpreted as an effect of destructive
  Aharonov-Bohm-type quantum interference between different paths of
  the tunneling electrons. Together with a blockade of some inelastic
  channels due to Pauli-principle restrictions (see text and the
  caption to Fig.~\ref{fig:inelastic}) this leads to a finite
  magnetoresistance of the wire. Amplitude shown is greatly
  exaggerated. Reprinted with permission from
  [\onlinecite{Aharanovpaper}], G. Sonne.}
\label{fig:Aharonov}
\end{figure}

Quantum fluctuation of the nanowire bending modes is a feature which
must be taken into account while considering the effects of quantum
coherence in the vibration dynamics of a suspended nanowire based
tunneling device (see Fig.~\ref{fig:Aharonov}).
This implies that geometrical constraint for the electrons, set by
wire geometry, will no longer localize them to the 1-dimensional
conducting wire, but will be also imply certain delocalization in the
direction transverse to the wire axis. Such change of the
dimensionality opens up the possibility for quantum interference
effects in the electronic tunneling over the leads if an external
magnetic field, $H$, is applied. As a result, a finite
magnetoconductance of the 1D vibrating wire occurs as a manifestation
of nanomechanical quantum coherence effects \cite{Shekhter2006}.

The voltage-biased suspended nanowire structure shown in
Fig.~\ref{fig:Aharonov} was first analyzed in
Ref.~[\onlinecite{Shekhter2006}] for the case of nonresonant electronic
transmission through the wire. Considering the electronic Fermi level
to be far from the energy levels of the quantized longitudinal
motion of the electrons in the wire, the authors used perturbation
theory with respect to the tunnel barrier transmission, $T_{eff}$, to
calculate the current and conductance through the system. To second
order in this expansion, the effective Hamiltonian coupling the
electrons in the two leads through the virtual energy level of the
wire can be shown to be of the form,
\begin{equation}
\label{Aharonovhamil}
\begin{split}
\hat{H}&=\sum_{\sigma,k}\varepsilon_{\sigma,k}\hat{a}^{\dagger}_{\sigma,k}\hat{a}_{\sigma,k}+\hbar\omega\hat{b}^{\dagger}\hat{b}\\*
&+e^{i\phi(\hat{b}^{\dagger}+\hat{b})}\sum_{k,k'}T_{eff}(k,k')\hat{a}_{r,k}^{\dagger}\hat{a}_{l,k'}+\text{h.c.}\,.
\end{split}
\end{equation}
Here, electrons in the leads are described by the first term while the
fundamental mode of the flexural vibrations\footnote{For simplicity,
  only the fundamental mode is considered in this analysis.} is
described by its harmonic oscillator Hamiltonian (the second term in
Eq.~\eqref{Aharonovhamil}). The additional phase factor
$e^{i\phi(\hat{b}^{\dagger}+\hat{b})}$ connected with the tunneling
electrons (third term in \eqref{Aharonovhamil}) describes the
magnetic field dependent phase which the electrons acquire between the
two sequential tunneling events at the ends of the wire.  The
specifics of this term for the problem considered is that the
electronic phase is not a c-number, but an operator acting in the
quantum vibrational space.

As the effective coupling strength for the tunneling electrons
experiences quantum fluctuations so does the number of different
tunneling channels. Since these channels obey quantum nanomechanical
dynamics, we obtain a typical picture of entanglement between the
quantum nanomechanical and electronic degrees of freedom.  The outcome
of this is that new polaronic states (referred to as ``swinging
states'' in Ref.~[\onlinecite{Shekhter2006}]) are formed in the wire
as a result of this entanglement, and charge transport can now be
viewed as electronic transmission through these intermediate
states. It is interesting to note that the strength of such polaronic
coupling is determined by magnetic flux and can thus be tuned by the
external magnetic field.

From the effective tunneling Hamiltonian, Eq.~\eqref{Aharonovhamil},
the electric current over the voltage-biased leads can easily be
calculated. What is new in this analysis is that the matrix elements
corresponding to electronic tunneling are now described through
operators in the vibrational space of the wire, something that needs
to be considered when tracing out the electronic and vibrational phase
space. The resulting expression for the current is presented in
Eq.~\eqref{Aharonovcurr},
\begin{align}\label{Aharonovcurr}
&I=\frac{G_0}{e}\sum_{n=0}^{\infty}\sum_{\ell=-n}^{\infty}P(n)\vert\langle n\vert e^{i(\hat{b}^{\dagger}+\hat{b}^{\dagger})}\vert n+\ell\rangle\vert^2\\*
&\times\int\textrm{d}\epsilon[f_l(\epsilon)(1-f_r(\epsilon-\ell\hbar\omega))-f_r(\epsilon)(1-f_l(\epsilon-\ell\hbar\omega))]\nonumber\,. 
\end{align}
Here, $G_0$ is the zero field conductance, $e$ is the electric charge
and the oscillating nanowire is assumed to be in thermal equilibrium
with the environment, described through
$P(n)=(1-e^{-\beta\hbar\omega})e^{-n\beta\hbar\omega}$, the
probability of finding the oscillator in a quantum state $n$ with
energy $n\hbar\omega$ at a temperature $T$ ($\beta=(k_BT)^{-1}$ and
$\omega$ is the frequency of oscillation of the nanowire). The
electron-vibron coupling is expressed through the coupling constant
$\phi$, which scales linearly with the magnetic field. Finally
$f_{l,r}$ are the Fermi functions for the left and right lead kept at
chemical potential $\mu_{l,r}=\pm eV/2$ respectively with $V$ the bias
voltage.
\begin{figure}
\includegraphics[width=0.25\textwidth]{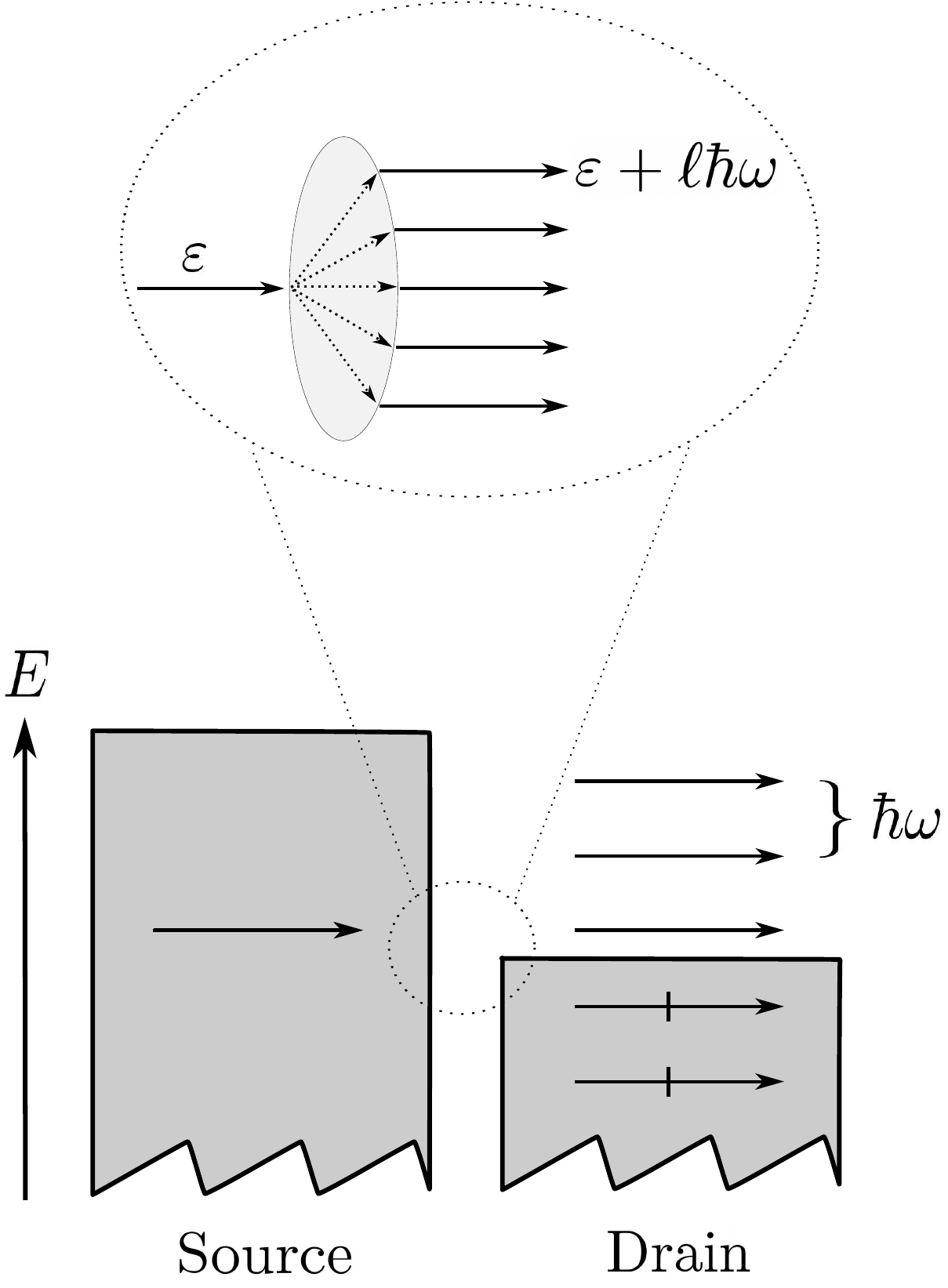}
\caption{Sketch of the different transmission channels available for
  electrons tunneling through the oscillating nanowire of
  Fig.~\ref{fig:Aharonov}. Electrons with energy $\varepsilon$
  tunneling from the left (source) to the right (drain) lead are
  transmitted in both elastic and inelastic tunneling channels (top
  image) with the corresponding energy exchange, $\varepsilon +
  \ell\hbar\omega$; $\ell=0, \pm 1, \pm 2,\ldots$.  Due to
  Pauli-principle restrictions, some of the inelastic channels are
  affected by the electronic population in the drain lead (shaded
  region, lower image), which together with a reduction of the
  tunneling rate in the elastic channel $\ell =0$ (see text and
  caption to Fig.~\ref{fig:Aharonov}) leads to a finite
  magnetoresistance of the wire. The Pauli-principle restrictions are
  important only if $\varepsilon$ is close to the chemical potential
  in the drain lead. This is why the total current reduction saturates
  and becomes independent of both temperature and bias voltage for
  large enough bias voltage (see text).}
 \label{fig:inelastic}
\end{figure}

The above expression represents contributions to the current from a
number of different inelastic tunneling channels, $\ell\neq 0$,
together with the elastic channel. Analyzing Eq.~\eqref{Aharonovcurr}
one finds that the destructive Aharonov-Bohm interference to the
elastic channel caused by the quantum fluctuations in the magnetic
flux (proportional to the area enclosed by the vibrating quantum wire)
is compensated by the additional inelastic tunneling channels. These
inelastic channels increase the electronic transport through the
system, and hence compensates the effect of the quantum suppression in
the elastic channel. The influence of these two effects on the total
transport depends on the extent to which the inelastic channels are
permitted by the Pauli principle, which limits the electron-vibron
energy transfer. The corresponding energy diagram for the different
electronic tunneling channels is presented in
Fig.~\ref{fig:inelastic}.

\begin{figure}
\begin{center}
\subfigure[]{\label{fig:Aharocondu}\includegraphics[width=0.4\textwidth]{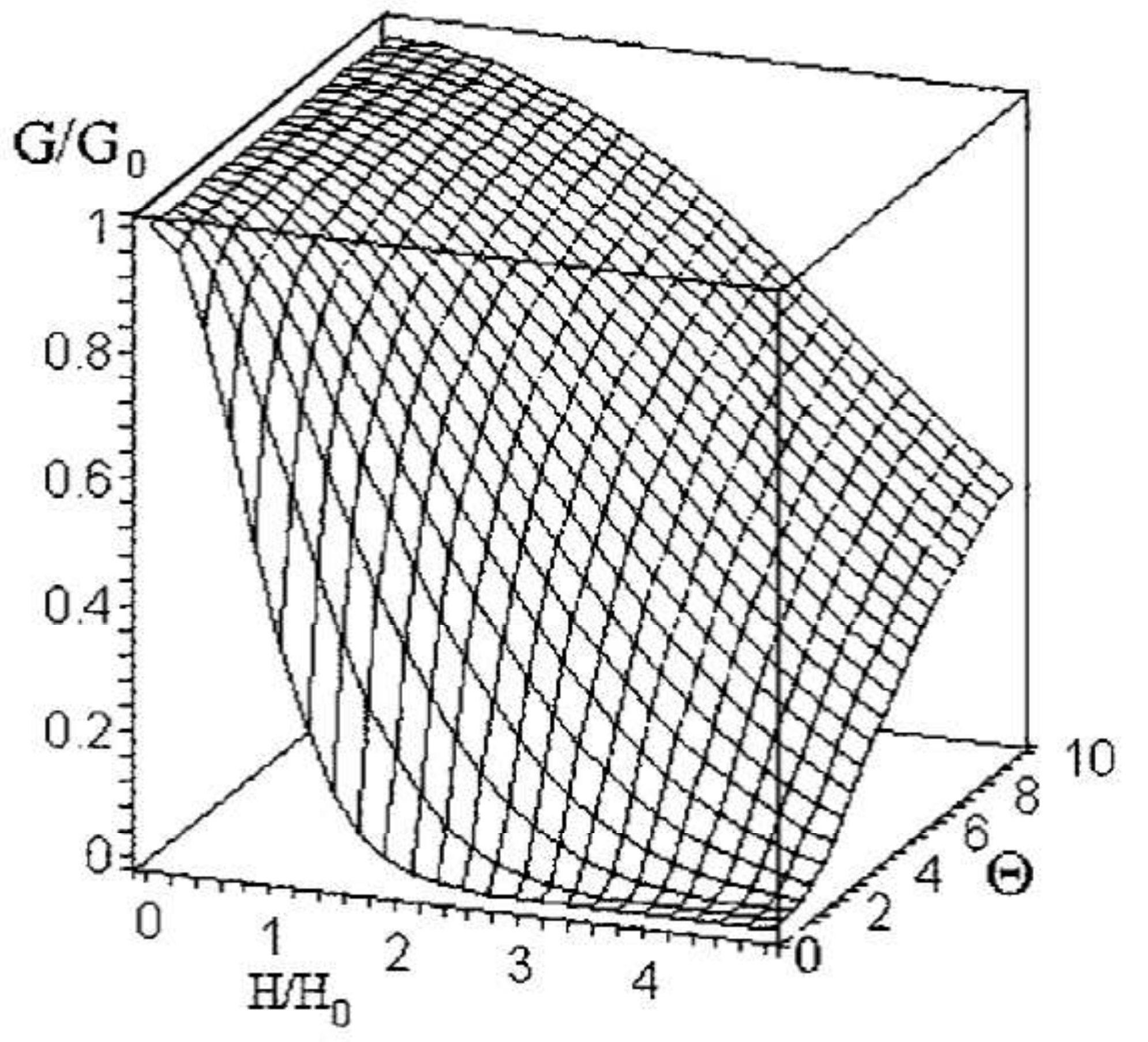}}
\subfigure[]{\label{fig:Aharanovcurrent}\includegraphics[width=0.45\textwidth]{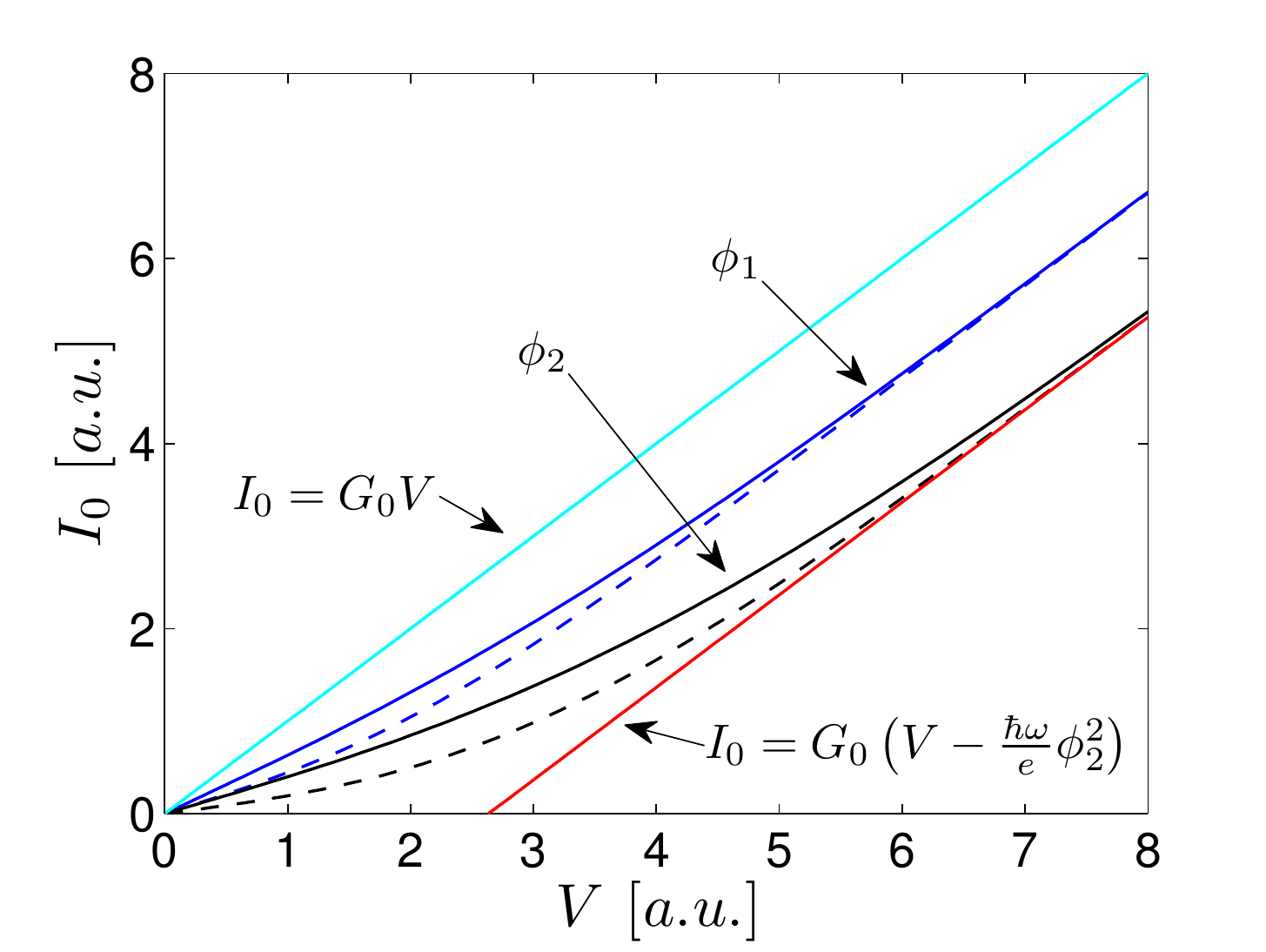}}
\caption{(Color online) \textbf{(a)} Magnetoconductance, $G/G_0$,
  through suspended SWNT system as a function of temperature
  $\Theta=k_BT/\hbar\omega$ and magnetic field $H/H_0$. Reprinted with
  permission from [\onlinecite{Shekhter2006}], R. I. Shekhter
  \textit{et al}, \textit{Phys. Rev. Lett}, \textbf{97}, 156801
  (2006). $\copyright$ 2006, American Physical Society. \textbf{(b)}
  Predicted offset current as a function of bias voltage for two
  different temperatures and two different vibrational
  frequencies. Note that the offset current (red line) does not
  extrapolate to the origin and scales as the square of the magnetic
  field, $\phi^2\propto H^2$. Data shown in arbitrary units. Reprinted
  with permission from [\onlinecite{Aharanovpaper}], G. Sonne.}
\end{center}
\end{figure}

Neglecting with the Fermi statistics of the electrons (high
temperature limit), the restrictions imposed by the summation over the
possible inelastic tunneling channels in Eq.~\eqref{Aharonovcurr}
disappears, and the expression for the conductance exactly coincides
with the transmission through the non-vibrating wire (see
Eq.~\eqref{Aharocondu}).  In the low temperature limit however, the
result of the abovementioned quantum mechanical suppression of the
elastic channel and the Pauli restrictions on some of the inelastic
channels is the appearance of a finite magnetoconductance. As a
result, the strongest effects of the predicted quantum
magnetoconductance occurs in the low voltage, low temperature limit:
$\hbar\omega\gg e V,k_B T$. At higher energies, the Pauli restrictions
become less important, and these effect only give corrections to the
conductance as compared to the non-vibrating wire. Mathematically the
asymptotic limits to the magnetoconductance are found as,
\begin{equation}
\frac{G}{G_0}=\left\{\begin{array}{ll} e^{-\phi^2/2} &\beta\hbar\omega\gg 1\\1-\frac{\hbar\omega}{6 k_B T}\phi^2 &\beta\hbar\omega\ll 1\,,
\end{array}\right.
\label{Aharocondu}
\end{equation}
which can also be seen in Fig.~\ref{fig:Aharocondu}. Similar results
of quantum mechanical effects on the conductance have been reported
for the case of ballistic electron transport through a carbon nanotube
encapsulating a single movable fullerene scatterer \cite{peapodpaper}
(see also Ref.~[\onlinecite{Krive2008}] for a resonant tunneling
treatment of a similar system).

The effects of the quantum suppression of the low energy tunneling
processes outlined above are only visible in the low temperature, low
bias voltage limit and are hence hard to verify experimentally. If,
however, the electrical current is instead considered these effects
are indeed visible also at higher temperatures and voltages. This can
be understood as the Pauli restrictions on the inelastic tunneling
channels affect only the low energy electrons (low bias
voltage). Thus, the low voltage current is reduced from the
non-vibrating ohmic current, $I=G_0V$, by an amount that is given by
the extent to which the elastic channel is suppressed (which in turn
depends on the magnetic field). As the bias voltage is increased the
inelastic tunneling channels are opened and the current increases
accordingly. However, a further increase in the bias voltage when the
conditions $eV\gg \ell\hbar\omega, k_BT$ are fulfilled will not be
affected by the Pauli restrictions due to the large energy scales of
the electrons. As a result, these effects are displayed as a
temperature- and bias voltage-independent current deficit, as compared
to the non-vibrating wire, as recently reported in
Ref.~[\onlinecite{Aharanovpaper}],
\begin{equation}
\lim_{eV\gg \ell\hbar\omega, k_B T}I=G_0\left(V-\frac{\hbar\omega}{e}\phi^2\right)\,.
\label{Aharodef}
\end{equation}
The effect of the quadratic magnetic field dependence on the current
deficit can be experimentally observed by extrapolating the high
voltage asymptotes of the $I-V$ curves to the $V\rightarrow 0$ limit
(see Fig.~\ref{fig:Aharanovcurrent}). For realistic parameters, we
estimate that these effects should be observable for bias voltages
$V\sim$~\unit[30]{$\mu$V} and currents $I\sim$~\unit[3]{pA} in
magnetic fields of $H\sim$~\unit[20]{T}.
\\\\

By exploiting the equilibrium distribution of the flexural vibrations,
$P(n)$ in Eq.~\eqref{Aharonovcurr}, we have neglected with the
non-equilibrium effects which can be stimulated in the vibronic
subspace due to the coupling to the current carrying electrons.  This
assumption is valid if the internal relaxation of the vibronic
subsystem is strong compared to the excitation strength set by the
electronic emission of vibrons. If this is not the case, such
electron-vibron coupling cannot be ignored, even if the electronic
tunneling is small.  In order to account for the mutual
electron-vibron dynamics, one should evaluate the evolution of the
density matrix operating on both the vibronic and electronic
subspaces, and, using this density matrix, calculate the vibronic
contribution to the current as well as the electron-vibron
contribution to the Lorentz force.  This analysis was recently
performed in Ref.~[\onlinecite{Aharanovpaper}], where it was shown
that the stationary density matrix can be found as a result of a
balance between the electron assisted emission and absorption of
vibrons. The equation for the stationary reduced density matrix,
$\hat{\rho}$, of the electron-vibron subspace takes the form,
\begin{align}\label{Aharodensity}
\frac{i}{\hbar}[&\hat{H}_{osc},\hat{\rho}]=\vert T_{eff}\vert^2\left[(\hat{J}_1+\hat{J}_2)\hat{\rho}+\hat{\rho}(\hat{J}_1^{\dagger}+\hat{J}_2^{\dagger})\right.\\*
&-\left.e^{-i \chi \hat{x}}(\hat{J}_1\hat{\rho}+\hat{\rho} \hat{J}_1^{\dagger})e^{i \chi \hat{x}}-e^{i \chi \hat{x}}(\hat{J}_2\hat{\rho}+\hat{\rho} \hat{J}_2^{\dagger})e^{-i \chi \hat{x}}\right]\notag\,.
\end{align}
Here, $\hat{J}_{1,2}$ are operators that take into account the
electron-vibron coupling and $\hat{H}_0$ is the first two terms in
Eq.~\eqref{Aharonovhamil}. Also, the average Lorentz force
\eqref{expectx} and momentum \eqref{expectp} on the wire can be found
by multiplying Eq.~\eqref{Aharodensity} with the deflection and
momentum operators and tracing out the nanowire's degrees of freedom,
\begin{subequations}\label{Aharaoexpect}
\begin{gather}
\label{expectx}\frac{\hbar\omega}{x_0^2}\langle \hat{x}\rangle =\vert T_{eff}\vert^2\textrm{Tr}\left((\hat{J}_1-\hat{J}_2)\hat{\rho}+\hat{\rho}(\hat{J}_1^{\dagger}-\hat{J}_2^{\dagger})\right)\\*
\label{expectp}\langle \hat{p}\rangle=0\,.
\end{gather}
\end{subequations}
In Ref.~[\onlinecite{Aharanovpaper}] it was shown that although each
inelastic electronic tunneling channel (see Eq.~\eqref{Aharonovcurr})
is significantly renormalized by the non-equilibrium of the vibronic
subsystem, the total high voltage limit, $eV\gg k_BT,\hbar\omega$, to
the current is still given by Eq.~\eqref{Aharodef}. Furthermore, any
corrections to this expression from the thermal environment was shown
to decay exponentially in this limit for all relaxation strengths in
the vibronic subsystem.

\subsection{\label{sec:Super} Nanoelectromechanics of suspended nanowire superconducting weak link}

Superconducting ordering in the electronic subsystem qualitatively
changes the electromechanical coupling in the suspended nanowire based
NEM tunneling devices considered so far. If the electronic subsystem
of the leads in Fig.~\ref{fig:Aharonov} undergoes a superconducting
phase transition a new type of electronic coupling between the leads
occurs. This coupling allows for the possibility of coherent tunneling
of a pair of electrons (a Cooper pair), forming a superconducting
condensate of electrons in each of the two leads. As the process of
Cooper pair tunneling is a dissipationless ground state property of
the superconductors, a finite dc current over the leads will be
present even at zero bias voltage (the dc Josephson effect). If
instead a finite dc bias is applied an alternating Josephson current
is set up over the leads, representing the tunneling response of the
junction to the bias, whose zero average value guarantees that no
electric power is absorbed from the voltage source.  The frequency of
these current oscillations is set by the energy gained by a single
Cooper pair during its transition between the superconducting leads:
$h\nu=2eV$ (the ac Josephson effect).  

If now the tunneling Cooper pairs are coupled to the mechanical
vibrations of the suspended nanowire, an electromotive force is
induced due to the presence of the external magnetic field causing the
vibrations\footnote{The inductive coupling between flexural wire
  vibrations and the supercurrent flow in a SQUID loop was considered
  in Refs.~[\onlinecite{Blencowe2007,Buks2006,Buks2008}].}.  The work
done by this force on a tunneling Cooper pair renormalizes the total
energy gained by it and hence changes the frequency of the Josephson
oscillations. On the other hand, the mechanical vibrations of the
nanowire are affected by the Lorentz force which oscillates in time
due to the ac variations of the Josephson current in the wire. From
these considerations a set of equations determining the Josephson
frequency renormalization and the mechanical motion of the wire were
derived in Ref.~[\onlinecite{drivenoscillator}]. These equations were
found by considering the Andreev level formation in the
NEM-superconducting weak link in the low tunnel barrier transparency
limit (see also Appendix~\ref{Append:super}) and are fully consistent
with the qualitative picture describe above. Using the dimensionless
coordinates $Y$, $\epsilon$ and $\widetilde{\gamma}$ for the wire's
deflection coordinate, driving force and phenomenological damping
respectively these equation are,
\begin{subequations}\label{driveneom}
\begin{gather}
\ddot{Y}+\widetilde{\gamma}\dot{Y}+Y=\epsilon\sin(\varphi)\\*
\label{driveneomphase}\dot{\varphi}=\widetilde{V}-\dot{Y}\,.
\end{gather}
\end{subequations}
In \eqref{driveneom}, $\widetilde{V}$ is the dimensionless applied
bias voltage and the driving force $\epsilon\propto
H^2$. Eq.~\eqref{driveneom} is thus the equation of motion for the
deflection coordinate of the wire which is driven by a force
proportional to a term reminiscent of the ac Josephson current,
$\propto\sin(\widetilde{V}t)$. Note however, that as the wire moves in
the magnetic field it induces an electromotive force proportional to
the time rate of change of the deflection coordinate, which
counteracts the motion causing the vibration. This term is given by
the second argument in Eq.~\eqref{driveneomphase} (see
Ref.~[\onlinecite{drivenoscillator}] for details), and is responsible
for the characteristic resonant phenomena described below.

\begin{figure}
\begin{center}
\hspace*{-0.01in}
\subfigure[$\,\,\,\epsilon\sim\widetilde{\gamma}$]{\label{fig:Q100}\includegraphics[width=0.35\textwidth]{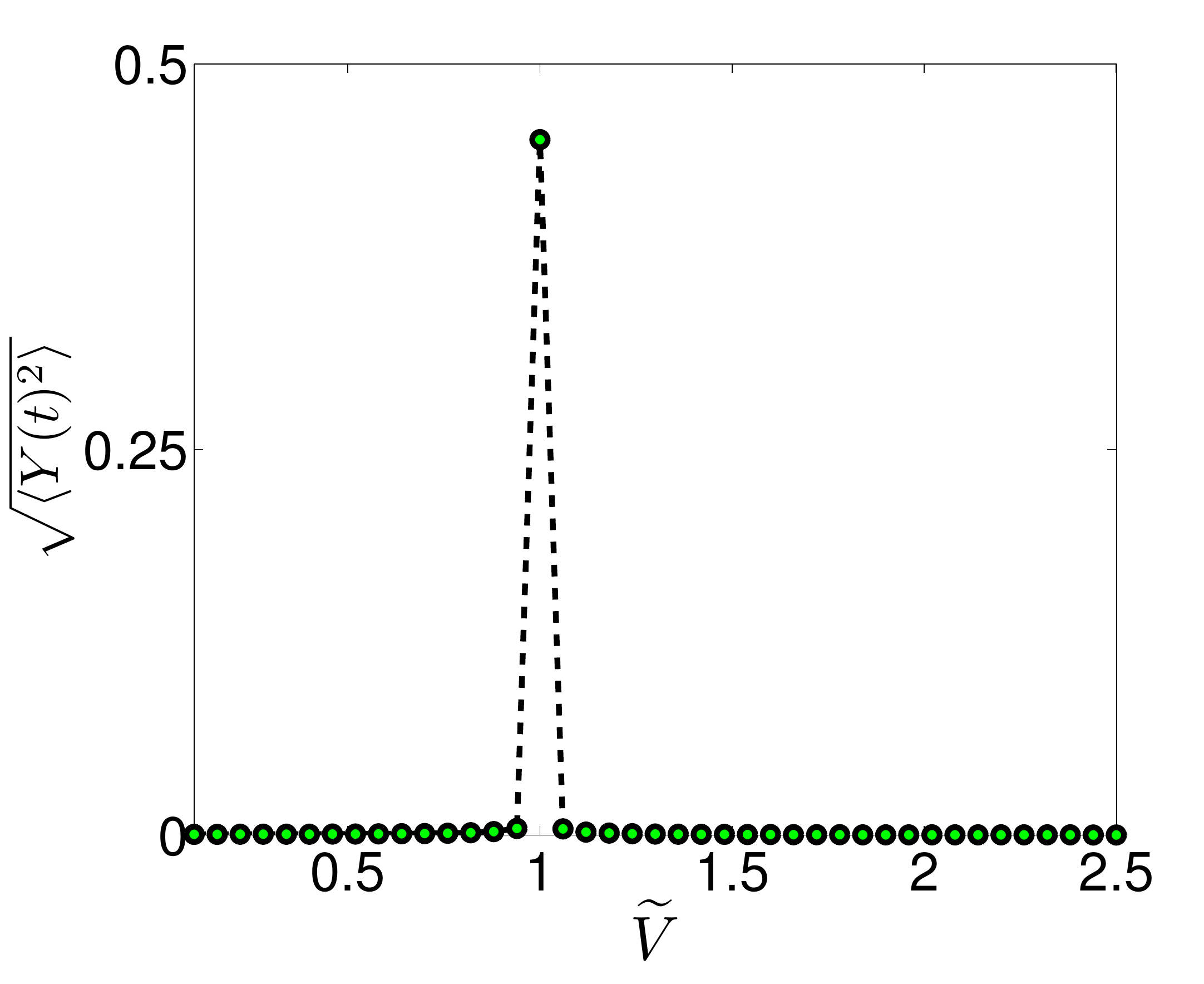}}
\subfigure[$\,\,\,\epsilon>\widetilde{\gamma}$]{\label{fig:Q1000}\includegraphics[width=0.35\textwidth]{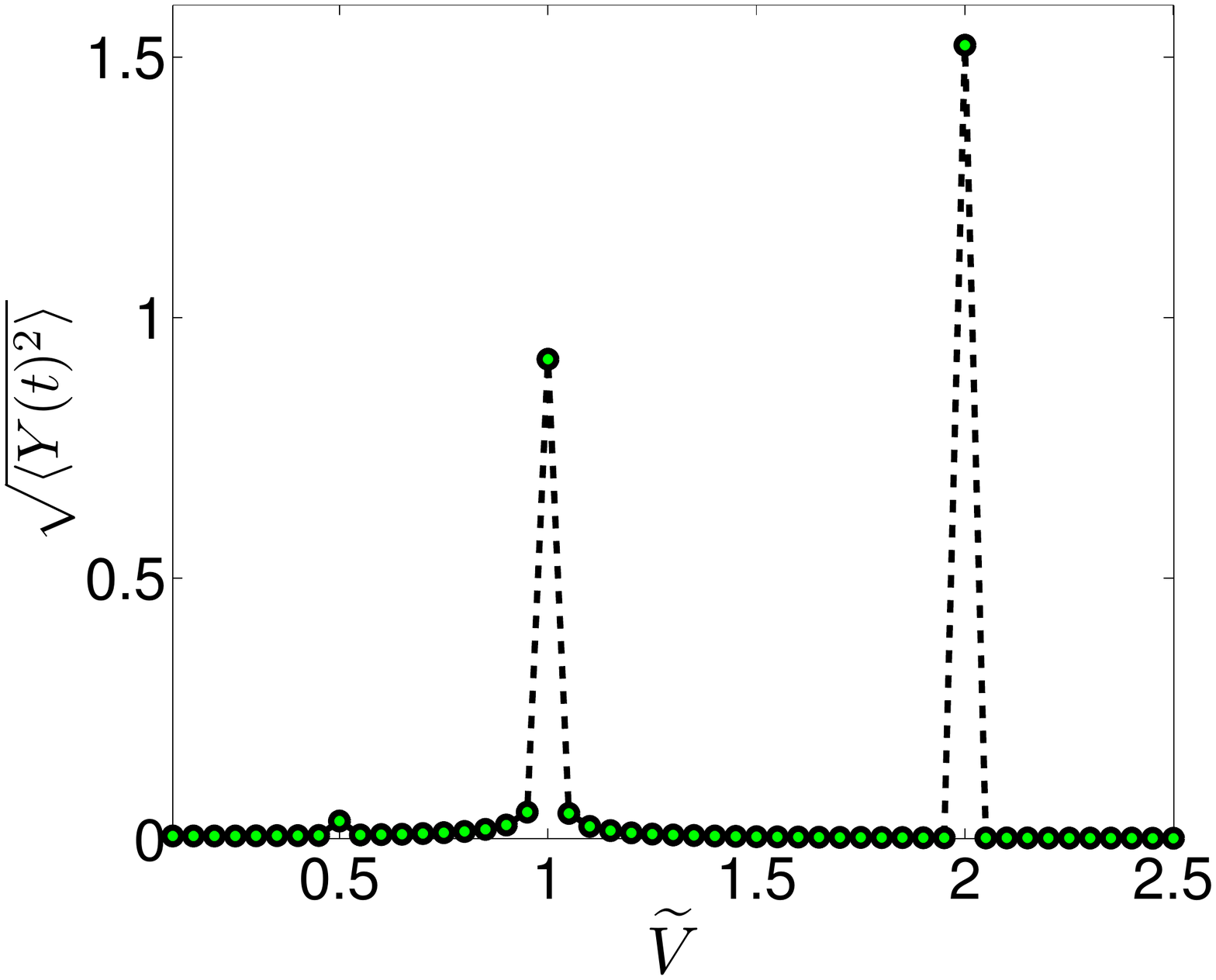}}
\caption{Time-averaged vibrational amplitude of the superconducting
  suspended nanotube as a function of driving voltage,
  $\widetilde{V}$, clearly showing the onset of the parametric
  resonance at higher driving force, $\epsilon$. Reprinted with
  permission from [\onlinecite{drivenoscillator}], G. Sonne \textit{et
    al.}, \textit{Phys. Rev. B}, \textbf{78}, 144501
  (2008). $\copyright$ 2008, American Physical Society.}
\label{fig:resonances}
\end{center}
\end{figure}

\begin{figure}
\begin{center}
\subfigure[]{\label{fig:eplot}\includegraphics[width=0.35\textwidth]{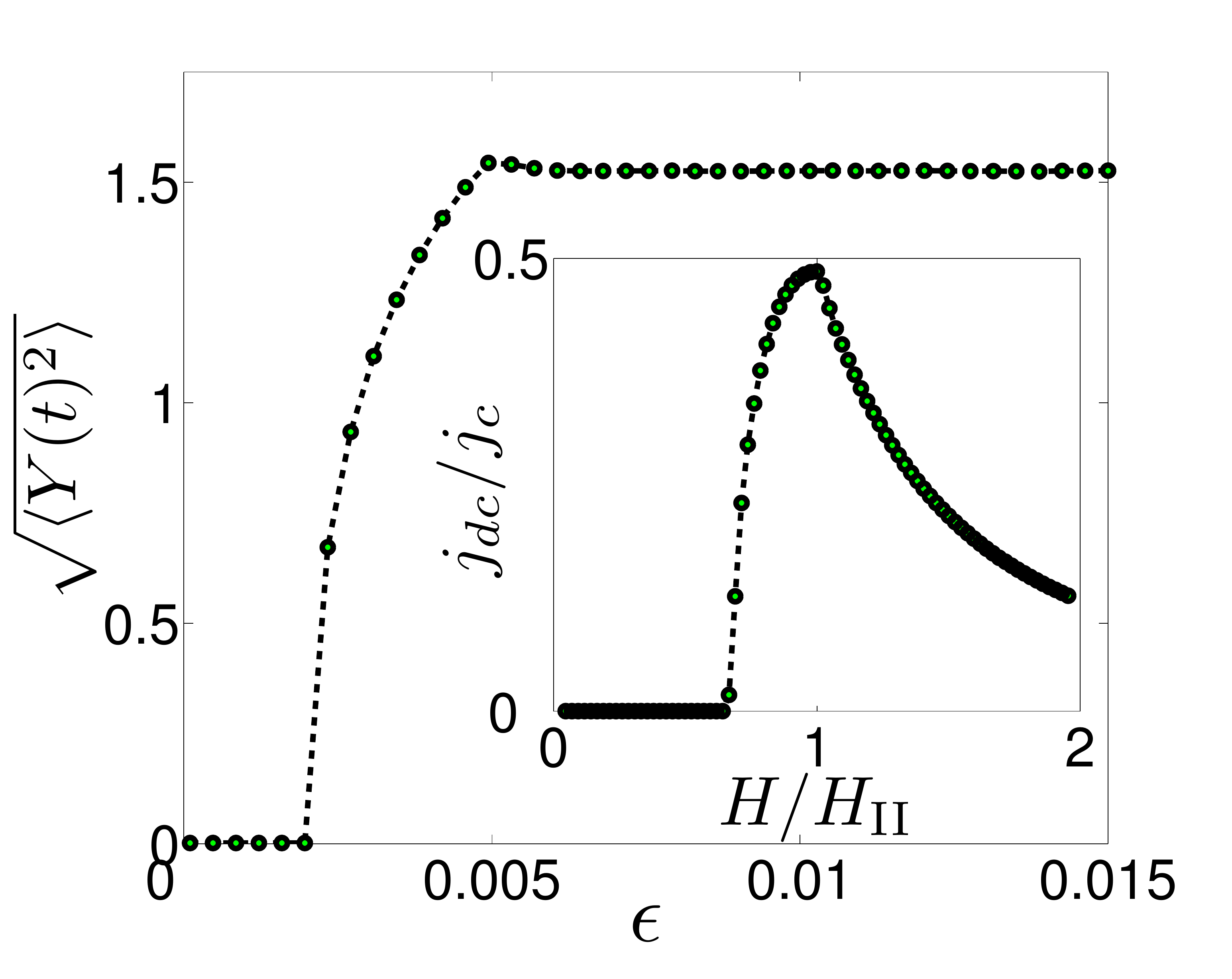}}
\subfigure[]{\label{fig:currenthyst}\includegraphics[width=0.39\textwidth]{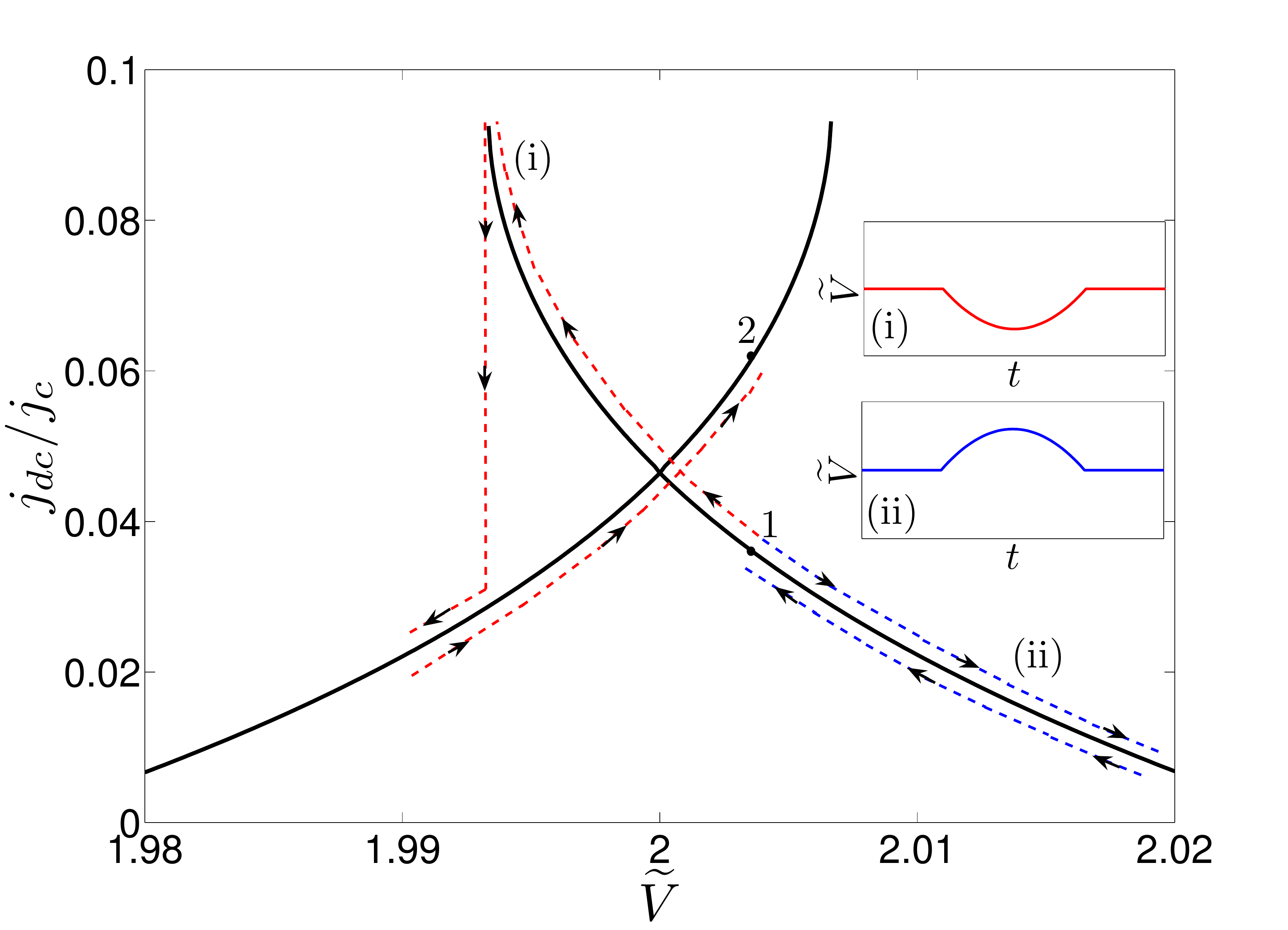}}
\caption{(Color online) \textbf{(a)} Time-averaged vibrational
  amplitude as a function of driving force for the second resonance
  peak, $\widetilde{V}=2$, clearly showing the saturation of the
  vibrational amplitude at $\epsilon^{*}\sim 0.005$. Inset shows
  corresponding dimensionless dc current as a function of the magnetic
  field. \textbf{(b)} Predicted current bistability and hysteresis for
  two voltage pulses close to the parametric resonance. As shown,
  applying either a positive or negative voltage pulse in time can
  shift the system between the two stability points (1 and 2) which
  are separated by a finite current difference. Reprinted with
  permission from [\onlinecite{drivenoscillator}], G. Sonne \textit{et
    al.}, \textit{Phys. Rev. B}, \textbf{78}, 144501
  (2008). $\copyright$ 2008, American Physical Society.}
\end{center}
\end{figure}

Numerical simulations of Eq.~\eqref{driveneom} shows that for small
driving forces (low magnetic fields) the system will achieve resonant
conditions only at the eigenfrequency of the fundamental mode of the
wire $\widetilde{V}=1$, see Fig.~\ref{fig:Q100}. For larger driving
forces however, also parametric excitations $\widetilde{V}=2$ can be
achieved, Fig.~\ref{fig:Q1000}.
Furthermore, the amplitude of these modes is initially seen to be an
increasing function of the magnetic field strength, but eventually
saturates at some critical driving force $\epsilon^{*}$,
Fig.~\ref{fig:eplot}. To explain these phenomena we derive two
stability equations for the amplitude and phase of the vibrating
nanowire which well describe the onset and saturation of finite
vibrations at resonant conditions. These equations also show that for
the system considered there exists a window of bistability in the
vibrational amplitude around the resonance peaks.

Being that the voltage source pumps energy into the oscillating
nanomechanical system, it can be shown by direct integration of the
equation of motion that at resonant conditions a finite dc current,
$j_{dc}$, is set up over the system. This comes about as the
mechanical energy of the nanowire is a constant of time at resonant
conditions, hence any pumping of the system at these conditions will
lead to an energy transfer through it, proportional to the real
damping coefficient $\gamma$. As such, the resonant dc current can be
shown to scale as,
\begin{equation}
j_{dc}\propto \frac{\gamma\langle \dot{Y}(t)^2\rangle}{L^2H^2 V}\,,
\label{suprecurrent}
\end{equation}
where time-average rate of change of the deflection coordinate,
$\langle \dot{Y}(t)^2\rangle$, behaves phenomenologically equivalent
to the deflection coordinate. As such, we predict that the system
considered should exhibit both negative and positive magnetoresistance
as the current is first an increasing function of the magnetic field
(increasing vibrational amplitude) but falls off as $j_{dc}\propto
H^{-2}$ once the driving force reaches the critical value
$\epsilon^{*}$ which indicates the onset of the bistability in the
vibrational amplitude (see inset Fig.~\ref{fig:eplot}).

Also, we have shown that the bistability region around the resonance
peaks can be utilized to directly probe the underlying quantum
mechanics of the system. As was mentioned above, close to resonance
conditions the system exhibits a mechanical bistability, i.e. two
dynamically stable solutions of oscillation can be found for the same
bias voltage but with different vibrational amplitudes. Which of the
two stability point the system will correspond to will in general
depend on initial conditions, however, by applying finite voltage
pulses in time one can move the system between the two. Consider for
example the situation shown in Fig.~\ref{fig:currenthyst} where the
system is initially found at the stable point 1, corresponding to the
lower vibrational amplitude (lower current) and the bias voltage is
slightly off the parametric resonance. Applying two voltage pulses in
time, (i) and (ii), will move the system along the phase space
trajectory of the stability point defining it. As the region of
bistability is only found within a small window of bias voltage around
the resonance peaks, pulse (i) will in this example force the system
into the second stability point, which will return the system to the
higher stability point, 2, on the up-sweep. As the dc current scales
directly with the vibrational amplitude we thus predict that this
pulse will result in a measurable current difference to pulse (ii)
which will return the system to the initial stability
point. Considering realistic parameters we predict that the measurable
current difference ($1\rightarrow 2$) $\sim$~\unit[5]{nA} should be
experimentally observable for magnetic fields $H\sim$~\unit[20]{mT}
and bias voltages $V\sim$~\unit[5]{$\mu$V}.

\section{Conclusions}

Coupling between electrical and mechanical degrees of freedom is the
basic mechanism behind the functionality of any
nanoelectromechanical system. Such coupling can be achieved in many
different ways. In this Review we have focused on the effects of
varying the spatial concentration of the electronic charge or
current in systems incorporating suspended nanowires as the
mechanical element. These structures are particularly interesting to
study as they can be used as efficient electromechanical transducers
that serve simultaneously as electric weak links for the tunneling
electrons and as nanomechanical resonators. This allows one to use
the mutual coupling between the electric current oscillations in
time and the mechanical vibrations to achieve a highly nonlinear
electromechanical coupling if strong driving voltages or
magnetic fields are applied. The resulting self-supported
nanoelectromechanical dynamics can be maintained through the
development of an electromechanical instability, caused either by
the shuttle phenomenon (for systems with varying charge
concentration) or through resonant pumping of nanovibrations by the
electric current flow (in systems with strong current
concentration).

In many of the systems considered in this Review, quantum coherence is
an essential feature of the discussed nanoelectromechanical
operations. For example, we have shown that quantum effects in the
electronic subsystem, such as coherent electron transmission through a
double barrier structure, can be coupled to the quantum coherence of
the mechanical flexural vibrations (Section~\ref{sec:Aharonov}). The
resulting electromechanical entanglement establishes ``swinging"
polaronic states for the tunneling electrons, which allows the quantum
vibrations to be investigated by interferometry. Also, we have shown
that the existence of more than one mode of nanomechanical bending of
the suspended nanowire, a direct consequence of the spatial extension
of the mechanically vibrating element, makes the operation of the NEM
device even richer. As an example we mention the multimode shuttle
instability discussed in Section~\ref{subsec:STM_displacement} where
it was shown that by applying a ``local'' external probe in the form
of current injection from an STM tip the ``global'' mechanical
properties of the system as a whole can be altered. In particular, the
competition between the different mechanical modes of vibrations will
lead to a self-organization of the shuttle vibrations of the suspended
nanowire.

Experimental realization of the phenomena discussed in this Review is
an experimentally challenging task. Although NEM-structures based on
suspended nanowires have been manufactured by a number of labs
\cite{Etaki2008,Huttel2008,Poncharal1999,Poot2007,Sapmaz2003,Sapmaz2006,Sapmaz2006-2,Witkamp2006,LeRoy2004,LeRoy2004a,Koenig2008},
the need for strong external magnetic fields and applied voltages
offers a real experimental challenge. Another obstacle to
experimental realization of these systems is damping of the
nanomechanical vibrations. If, for example, the mechanical damping is
too large, then the voltage threshold required for the onset of the
electromechanical instability may well exceed the thermal limit, set
by Joule heating of the system, and the system might burn
\cite{Johannes}. However, for the NEM systems considered here typical
experimental parameters such as quality factors of the order $Q\sim
1000$ and bias voltages of order \unit[1]{V} the suggested multimode
shuttling and superconducting pumping of nanovibrations should be
experimentally observable. The effects of the interferometry of the
quantum nanovibrations discussed here do nevertheless demand the
presently best achievable experimental conditions.

In this Review we have discussed the coupling of mesoscopic electrons
to nanomechanical radio frequency vibrations. It is interesting to
compare this to how mesoscopic electrons couple to external
electromagnetic fields of much higher frequencies. Such coupling has
been been shown to result in a number of interesting non-equilibrium
mesoscopic phenomena in both normal \cite{Grincwajg1995,Gorelik1997}
and superconducting \cite{Gorelik1995,Gorelik1998a} mesoscopic
structures if the microwave photon energy is comparable to the
electronic energy-level spacing.  The much lower mechanical
frequencies of NEMS devices, on the other hand, can easily be tuned to
match the smaller electronic energy scales that characterize the
widths (due to tunneling) of electronic energy levels, the applied
bias voltage, or the frequency of Rabi oscillations in the population
of energy levels. The fact that electrons in mesoscopic systems can
couple to oscillations corresponding to such very different energy
scales, makes it tempting to speculate that mesoscopic electrons might
be used for making microwave-mechanical transducers by simultaneously
coupling them to both electromagnetic and nanomechanical degrees of
freedom.  If such an electron-mediated microwave-mechanical coupling is
possible in principle, one may wish to consider quantum coherence
effects in structures where both electromagnetic, mechanical and
electronic degrees of freedom have been quantum mechanically
entangled. The extent to which such speculations are realistic is an
open question which we leave for future research in the exciting field
of nanoelectromechanics to answer.

\vspace{0.5mm}{\bf Acknowledgments} This work was supported in parts
by the Swedish VR and SSF, by the Faculty of Science at the
University of Gothenburg through its ``Nanoparticle" Research
Platform and by the Korean WCU program funded by MEST through
KOSEF (R31-2008-000-10057-0).

\appendix
\section{\label{Append:super} Derivation of the expression for the force on the pumped nanomechanical system}
Consider the Hamiltonian, $\hat{\mathcal{H}}$, for the electronic
subsystem of the pumped Josephson vibrations of the suspended
nanotube, Eq. (1) in Ref.~[\onlinecite{drivenoscillator}],
\begin{gather}
\hat{\mathcal{H}}=\int \textrm{d}x\hat{\Psi}^{\dagger}(x)\left(\hat{\mathcal{H}}_0+\hat{\mathcal{H}}_\Delta\right)\hat{\Psi}(x)\notag\\
\hat{\mathcal{H}}_0=-\frac{\hbar^2}{2m}\sigma_z\left(\frac{\partial }{\partial x}-\sigma_z\frac{i e H u(x,t)}{\hbar}\right)^2+\sigma_z U(x)\notag\\
\hat{\mathcal{H}}_\Delta=\Delta(x)\left[\sigma_x\cos\phi(t)+\text{sgn}(x)\sigma_y\sin\phi(t)\right]\,.
\label{superhamil}
\end{gather}
In \eqref{superhamil}, $\hat{\Psi}^{\dagger}$ [$\hat{\Psi}$] are
two-component Nambu spinors, $\sigma_i$ are the Pauli matrices in
Nambu space and the deflection of the tube is given by
$u(x,t)=u(x)a(t)$, where $u(x)$ is the dimensionless, normalized
profile of the fundamental bending mode and $a(t)$ determines the
amplitude of vibration. The potential $U(x)$ describes the barrier
between the nanotube and the bulk superconducting electrodes where the
gap parameter is $\Delta(x)=\Delta_0\Theta(2\vert x\vert-L)$ and
$\phi(t)=2 e V t/\hbar$ is the phase difference across the
junction. As indicated in Ref.~[\onlinecite{drivenoscillator}] a gauge
transform, $e^{i\hat{S}}\hat{\mathcal{H}}e^{-i\hat{S}}$ with
$\hat{S}=\sigma_z e H\int_{0}^{x}u(x',t)\textrm{d}x'/\hbar$, shifts
the dependence of the vector potential induced by the nanotube
deflections from the kinetic part of the Hamiltonian to the phase
differences over the leads,
$\phi(t)\rightarrow\varphi(t)=\phi(t)-a(t)4e
H\int_0^{L/2}u(x)\textrm{d}x/\hbar$.

Suppose now that we have an eigenstate of the system,
$\vert\psi(t)\rangle$, such that
$\vert\psi(t)\rangle=e^{-i\int_0^tE_0(t')\textrm{d}t'/\hbar}(\vert\psi_0(t)\rangle+\vert\tilde{\psi}(t)\rangle)$
where $\vert\psi_0(t)\rangle$ is the ground state of the system
corresponding to energy $E_0(t)$ and $\vert\tilde{\psi}(t)\rangle$ is
some state orthogonal to the ground state. To be able to evaluate the
force on the system with respect to the ground state we need to show
that $\vert\tilde{\psi}(t)\rangle$ is small, which can be done by
expanding it in a complete set of eigenstates of the Hamiltonian,
$\vert\tilde{\psi}(t)\rangle=\sum_{n}A_n(t)\vert\psi_n(t)\rangle$. Inserting
this form into the Schrödinger equation and
multiplying from the left with the state $\vert \psi_n(t)\rangle$, we
find that the amplitudes $A_n(t)$ can be expressed as,
\begin{equation}
A_n(t)=\frac{i\hbar\langle \psi_n(t)\vert\partial_t\psi_0(t)\rangle}{E_n(t)-E_0(t)}\,.
\label{superamps}
\end{equation}
To evaluate this further we need to find and expression for the term
in the nominator of Eq.~\eqref{superamps} which can be done by
considering the time derivative of the eigenstate solution
$\hat{\mathcal{H}}\vert\psi_0(t)\rangle=E_0\vert\psi_0(t)\rangle$ and
evaluating the matrix elements of this equation with
$\vert\psi_n(t)\rangle$. Performing this analysis we find that we can
express Eq.~\eqref{superamps} as,
\begin{gather}
A_n(t)=-\frac{i\hbar\langle \psi_n(t)\vert(\partial_t\hat{\mathcal{H}})\psi_0(t)\rangle}{(E_n(t)-E_0(t))^2}\,.
\end{gather}
As the only time-dependence of the transformed Hamiltonian is found in
the phase difference over the leads we identify that the nominator of
this expression is proportional to the bias voltage,
$\partial_t\hat{\mathcal{H}}\propto eV$, whereas the denominator is
the spacing of the energy levels in the systems, $\propto
\Delta_0$. Hence, for the system considered $eV\ll \Delta_0$, we can
safely evaluate the expression for the force with respect to the
fixed-phase ground state $\vert\psi_0(t)\rangle$.

The force on the wire is then found by evaluating the expectation
value of the force operator $\hat{F}\propto
\partial\hat{\mathcal{H}}/\partial a$ which can easily be shown to be
$F=\langle\psi_0\vert\hat{F}\vert\psi_0\rangle=-\partial
E_0[\varphi(a)]/\partial a$, where $E_0(\varphi)$ is the ground state
energy as given in Ref.~[\onlinecite{drivenoscillator}]. Similarly the
Josephson current can be shown to be $j=(2e/\hbar)\partial
E_0(\varphi)/\partial \varphi$ and we recover the total expression for
the force on the oscillating wire as given in Eq.~\eqref{driveneom}.

\bibliography{Review}

\end{document}